\documentclass[journal]{IEEEtran}
\usepackage{amsmath,amsfonts}
\usepackage{algorithmic}
\usepackage{array}
\usepackage[colorlinks=true, linkcolor=blue, citecolor=blue, urlcolor=blue]{hyperref}
\usepackage{textcomp}
\usepackage{stfloats}
\usepackage{url}
\usepackage{listings}
\usepackage{verbatim}

\def\BibTeX{{\rm B\kern-.05em{\sc i\kern-.025em b}\kern-.08em
    T\kern-.1667em\lower.7ex\hbox{E}\kern-.125emX}}
\usepackage{balance}

\usepackage[T1]{fontenc}
\usepackage{inconsolata}
\usepackage{tcolorbox}
\tcbuselibrary{skins}
\usepackage{booktabs}
\usepackage{multirow}
\usepackage{makecell}
\usepackage{tabularx, hhline}
\usepackage{tikz}
\usepackage[table]{xcolor} 
\usepackage{amssymb}
\usepackage{varwidth}

\usepackage{graphicx}
\usepackage{subcaption}

\usepackage{mdframed}

\usepackage{xcolor}

\newtcolorbox{themebox}[1]{
  colback=gray!5, 
  colframe=gray!80, 
  fontupper=\bfseries, 
  arc=0mm, 
  boxrule=0.5pt, 
  left=2pt, right=2pt, top=0.1pt, bottom=0.1pt,
  sharp corners,
  enhanced,
  overlay={\draw[line width=2pt,gray!80] (frame.north west)--(frame.south west);}
}

\lstdefinestyle{ptmcode}{
  basicstyle=\ttfamily\scriptsize,
  columns=fullflexible,
  breaklines=true,
  frame=single,
  rulecolor=\color{black!20},
  keywordstyle=\color{black}\bfseries,
  commentstyle=\color{green!60},
  stringstyle=\color{red!70},
  showstringspaces=false,
  captionpos=t
}

\lstdefinestyle{no_rationale}{
  basicstyle=\ttfamily\scriptsize,
  columns=fullflexible,
  breaklines=true,
  frame=single,
  rulecolor=\color{black!20},
  showstringspaces=false,
  escapeinside={(*}{*)},
  captionpos=t
}

\newcounter{summary}

\newenvironment{summarybox}[1]{ 
\refstepcounter{summary}
\begin{tcolorbox}[
  colback=gray!20,
  colframe=black,
  boxrule=0.8pt,
  arc=3pt,
  top=1pt, bottom=1pt, left=5pt, right=5pt,
  title={\textbf{Summary of Findings (#1)}},
  fonttitle=\bfseries,
  coltitle=black,
  colbacktitle=gray!60,
  titlerule=0.8pt,
  toptitle=1pt,
  bottomtitle=1pt
]
}{
\end{tcolorbox}
}

\tcbuselibrary{skins}

\definecolor{colorAdd}{HTML}{2ca02c} 
\definecolor{colorRem}{HTML}{d62728}
\definecolor{colorMig}{HTML}{1f77b4}
\definecolor{colorUpd}{HTML}{F4D35E}

\newcommand{\themebar}[4]{%
    \begin{tikzpicture}[baseline=0.3ex]
        \pgfmathsetmacro{\totalwidth}{2.0} 
        \pgfmathsetmacro{\wAdd}{(#1/#4)*\totalwidth}
        \pgfmathsetmacro{\wRem}{(#2/#4)*\totalwidth}
        \pgfmathsetmacro{\wMig}{(#3/#4)*\totalwidth}
        
        \fill[gray!10] (0,0) rectangle (\totalwidth, 0.2);
        
        \fill[colorAdd] (0,0) rectangle (\wAdd, 0.2);
        \fill[colorRem] (\wAdd,0) rectangle (\wAdd+\wRem, 0.2);
        \fill[colorMig] (\wAdd+\wRem,0) rectangle (\wAdd+\wRem+\wMig, 0.2);
        
        \draw[gray!30, very thin] (0,0) rectangle (\totalwidth, 0.2);
    \end{tikzpicture}%
}

\newcommand{\depthemebar}[5]{%
    \begin{tikzpicture}[baseline=0.2ex]
        \pgfmathsetmacro{\totalwidth}{2.0} 
        \pgfmathsetmacro{\wAdd}{(#1/#5)*\totalwidth}
        \pgfmathsetmacro{\wRem}{(#2/#5)*\totalwidth}
        \pgfmathsetmacro{\wMig}{(#3/#5)*\totalwidth}
        \pgfmathsetmacro{\wUpd}{(#4/#5)*\totalwidth}
        
        \fill[gray!10] (0,0) rectangle (\totalwidth, 0.2);
        
        \fill[colorAdd] (0,0) rectangle (\wAdd, 0.2);
        \fill[colorRem] (\wAdd,0) rectangle (\wAdd+\wRem, 0.2);
        \fill[colorMig] (\wAdd+\wRem,0) rectangle (\wAdd+\wRem+\wMig, 0.2);
        \fill[colorUpd] (\wAdd+\wRem+\wMig,0) rectangle (\wAdd+\wRem+\wMig+\wUpd, 0.2);
        
        \draw[gray!30, very thin] (0,0) rectangle (\totalwidth, 0.2);
    \end{tikzpicture}%
}

\lstdefinestyle{tightquery}{
  basicstyle=\ttfamily\footnotesize,
  breaklines=true,
  breakatwhitespace=true,
  columns=fullflexible,
  keepspaces=true,
  showstringspaces=false,
  frame=single,
  xleftmargin=0pt,
  aboveskip=3pt, belowskip=3pt,
  postbreak=\mbox{\texttt{\tiny$\hookrightarrow$}}\space
}

\begin{document}
\title{When AI Models Become Dependencies: Studying the Evolution of Pre-Trained Model Reuse in Downstream Software Systems}
\author{Peerachai Banyongrakkul, Mansooreh Zahedi, Christoph Treude, Haoyu Gao, Patanamon Thongtanunam
\IEEEcompsocitemizethanks{%
\IEEEcompsocthanksitem Peerachai Banyongrakkul: pbanyongrakk@student.unimelb.edu.au
}}

\markboth{Journal of \LaTeX\ Class Files,~Vol.~18, No.~9, September~2020}%
{How to Use the IEEEtran \LaTeX \ Templates}

\maketitle

\begin{abstract}
Modern software systems have transitioned from purely code-based architectures to AI-integrated systems where pre-trained models (PTMs) serve as permanent dependencies. However, while the evolution of traditional software libraries is well-documented, we lack a clear understanding of how these ``PTM dependencies'' change over time. Unlike libraries, PTMs are characterized by opaque internals and less standardized, rapidly evolving release cycles. Furthermore, their multi-role nature enables developers to treat individual instances of a single PTM as separate functional dependencies based on their specific downstream tasks. This raises a critical question for software maintenance: do PTMs change like standard software libraries or do they follow a divergent pattern? To answer this, we present the first empirical study of downstream PTM changes, analyzing a comprehensive dataset of 4,988 releases across 323 GitHub OSS repositories that reuse open-source PTMs. Using traditional software libraries as a baseline, we find that PTMs follow a qualitatively distinct pattern. PTMs are typically added late in the project life-cycle and tend to accumulate rather than be replaced as a project matures. Our findings show that PTM changes are three times less frequent (406 of 2,814 release transitions) than library changes. PTM changes are also less routinely documented, but more likely to carry explicit rationale. Unlike libraries, which evolve reactively, PTM evolution is proactively driven by capability expansion, with a unique documented rationale of PTM testing uncertainty. Our work calls for a rethinking of how PTMs are tracked and managed as dependencies in modern software engineering.
\end{abstract}


\section{Introduction}
Software reuse is a cornerstone of modern software engineering, allowing developers to accelerate development and enhance reliability by leveraging existing components \cite{Frakes2005, Yeow2025}. While the maintenance and risks of traditional library dependencies are well-documented \cite{Kula2018, Decan2019}, the rise of Artificial Intelligence (AI) has introduced a transformative paradigm: ``the reuse of Pre-Trained Models (PTMs)''. Facilitated by upstream PTM hubs like Hugging Face, developers now integrate complex capabilities, such as those found in Gemma-4\footnote{The most recent PTM in the Gemma family is available via Hugging Face: \url{https://huggingface.co/blog/gemma4}}, through simple function calls like \texttt{from\_pretrained()}. This shift has effectively turned PTMs into a new class of \emph{model-centric dependencies} (i.e., reusable AI models that software systems depend on for specific functional capabilities, like traditional software libraries) \cite{Jiang2023, Han2021, Yasmin2025}.

Previous studies have investigated challenges in PTM adoption from a broad ecosystem perspective~\cite{Jiang2023, Taraghi2024, Tan2024}. For example, they identified performance inconsistencies and difficulties in interpreting PTM outputs, drawing on evidence from Hugging Face and Stack Overflow. Our prior work~\cite{banyongrakkul2025} shifted focus to downstream developers (i.e., developers who integrate and maintain PTMs within their own software systems) and provided the first empirical investigation of challenges introduced by PTM reuse in actual open-source software (OSS) projects on GitHub. Yet, integration marks only the beginning of a longer maintenance journey. As projects mature, developers must manage changes in PTMs by adding, removing, or migrating. While these actions resemble traditional library evolution on the surface, PTMs may introduce unique complexities. They encapsulate not only code but also trained weights, architectures, and data-dependent behaviors \cite{Jiang2023}. Furthermore, PTM reuse is characterized by rapid evolution \cite{Jones2024}, sparse documentation \cite{Gao24}, inconsistent versioning practices \cite{Ajibode2025}, and their use for multiple functional roles within projects \cite{Yasmin2025}.

Despite the rapid adoption of PTMs driven by the growing number of PTMs from upstream hubs~\cite{Jiang2024} and increasing contributor requests for PTM additions~\cite{banyongrakkul2025}, little is known about how downstream developers manage PTM changes over time. This gap matters because unmanaged growth of model-centric dependencies may increase maintenance overhead, making it harder to track PTM usage, assess impact of changes and address security or licensing risks \cite{Davis2023}. Recent perspectives on AI-enabled software systems also suggest that software health may depend not only on code quality, but also on maintaining system understanding and documenting change rationale over time \cite{storey2026}. Crucially, it remains unknown whether PTM evolution patterns---how often PTMs change (i.e., additions, removals, and migrations) and why developers change them---mirror or diverge from traditional library practices.

To address this gap, we conducted the first comparative study of PTM evolution in downstream OSS repositories. We designed our data collection pipeline, inspired by PeaTMOSS \cite{Jiang2024}, to capture PTM reuse with significantly higher coverage and precision. Our methodology employed large-scale code search and rigorous validation of PTM signatures to curate a dataset of 4,988 releases across 323 PTM-enabled projects, allowing us to analyze PTM change patterns. We developed heuristics to identify PTM dependency changes, including additions, removals, and migrations. Through both quantitative and qualitative analyses, we examined the frequency, documentation practices, and the rationales developers document to justify PTM changes, in direct comparison to traditional software libraries.

Our findings reveal that PTM changes occur three times less frequently than library changes and follow an accumulative, addition-dominated pattern as projects mature. Although PTM changes are documented less frequently than library changes, they more often include explicit rationales, particularly for migrations. Qualitatively, we identify five primary documented rationales for PTM changes: new functionality expansion, insufficient PTM testing, performance pressure, upstream changes, and data pipeline gaps. While the expansion of new functionality is the leading driver, we uncover a unique PTM-specific driver regarding insufficient PTM evaluation and testing safeguards. In these cases, developers introduce additional PTMs specifically to benchmark or validate existing ones, rather than removing or replacing them to maintain ecosystem compatibility as is common with libraries.

Taken together, these results suggest that PTM reuse is not governed by replacement and upgrade cycles, but instead follows a pattern of accumulation driven by expanding functionality and PTM uncertainty. This accumulation could increase the number of coexisting PTMs and their associated dependencies, tests, and configurations, expanding the system’s maintenance surface over time. As a result, projects could face increased management complexity, such as tracking active PTMs, maintaining consistency across pipelines, and identifying stale or unused PTMs---all of which may contribute to technical debt if left unaddressed. These observations highlight several implications for PTM maintenance across three stakeholder groups: researchers can explore portfolio-oriented analysis and tooling to manage accumulation complexity; downstream developers could move toward more structured management of PTM usage and roles; and upstream hub providers can facilitate this by standardizing PTM versioning and validation infrastructure. Together, these efforts would help bridge the gap between traditional software engineering and the unique lifecycle requirements of PTM-integrated systems, reducing the risk of technical debt accumulation over time.

The key contributions of our research are as follows:
\begin{itemize}
    \item Curating and sharing a comprehensive dataset of 4,988 releases across 323 downstream Github OSS projects reusing PTMs, enabling the first longitudinal study of PTM change patterns in the OSS ecosystem\footnote{\label{fn:replication}Our replication package, including the dataset, scripts, analysis artifacts, and the online appendix, is available at:~\url{https://github.com/jaypeerachai/TSE_PTM_changes}}.
    \item Proposing a heuristic approach to detect PTM changes across release snapshots, enabling the first empirical study of how PTMs are added, removed, and migrated across downstream OSS projects.
    \item Deriving a three-level taxonomy of documented PTM change rationales through qualitative analysis of 209 documentation artifacts, comprising 70 codes, 20 sub-themes, and five high-level themes, including PTM-specific rationales.
    \item Constructing a cross-domain comparison with third-party library changes, revealing differences in change patterns, documentation, and underlying rationales.
\end{itemize}

The remainder of this paper is organized as follows: Section~\ref{sect:related_work} reviews related work and highlights research gaps; Section~\ref{sect:rqs} states our research questions; Section~\ref{sect:dataset} describes our data collection; Section~\ref{sect:method} illustrates the methodological framework used to answer our research questions; Sections~\ref{sect:result1} and \ref{sect:result2} present our results; Section~\ref{sect:discussions} provides a discussion of our findings and their implications; Section~\ref{sect:threats} discusses threats to validity; and Section~\ref{sect:conclusion} concludes the study.

\section{Background and Related Work}
\label{sect:related_work}

\subsection{Software Reuse and Dependency Management}

\emph{Software reuse} \cite{Frakes2005} refers to incorporating existing software artifacts into new systems to improve efficiency and reduce cost \cite{Yeow2025}.  It takes many forms, from source code reuse through object-oriented inheritance \cite{Giordano2023} and design reuse through patterns \cite{Gamma1993}, to reliance on external components. When a project relies on an external component, it introduces a \emph{dependency}. The most prevalent form is \emph{third-party library reuse}, integrating externally developed code managed through package managers such as Maven, NPM, and PyPI~\cite{Decan2019}. While this improves productivity \cite{Yeow2025}, it also introduces maintenance risks \cite{Cox2019}, and many projects lag behind in updating due to concerns about breaking changes \cite{Kula2018, Zerouali2018, Huang2022, Bavota2015, Pinckney2023}. In this paper, we define a \textit{change} as any modification to a project's set of external dependencies, including additions, removals, migrations, and version updates.

Given these challenges, several studies \cite{Teyton2014, Zaimi2015, Kabinna2016, He2021, Barbosa2022, Terzi2022, Jaisri2025} have investigated how frequently and in what ways developers actually change their library dependencies. These changes are typically detected by comparing dependency declaration files across project snapshots, such as \texttt{pom.xml} for Java \cite{He2021, Huang2022} or \texttt{requirements.txt} for Python \cite{Islam2023}. Overall, changes are infrequent. Only 5.57\% to 28.72\% of projects exhibited at least one library change \cite{Teyton2014, Zaimi2015, He2021}. They also observed that the average addition rate ranges from only 3\% to 9\% over project evolution \cite{Zaimi2015}. Migration rates are highly context-dependent across ecosystems \cite{He2021, Gu2023}, with notably higher rates in specific domains such as 14\% for logging \cite{Kabinna2016} and 66\% for testing libraries \cite{Barbosa2022}. In contrast, removals occur more frequently. He et al. \cite{He2021} and Jaisri et al. \cite{Jaisri2025} reported that 30.44\% to 41.04\% of projects removed at least one library over their history. Notably, library changes show no correlation with overall project growth \cite{Terzi2022}.

Several studies have explored why developers delay updating libraries, citing extra effort \cite{Kula2018}, frequent upstream updates \cite{Zerouali2018}, and breaking change concerns \cite{Bavota2015, Pashchenko2020}. When changes do occur, migrations are primarily motivated by performance, compatibility, or configuration concerns \cite{Teyton2014, Kabinna2016}, as well as maintenance, usability, and security needs \cite{He2021, Barbosa2022}. Removal decisions are commonly driven by library heaviness, redundant functionality, or security concerns \cite{Pashchenko2020, Jaisri2025}. Beyond triggers, Islam et al.~\cite{Islam2024} found that while most library migrations are small in scope, many require additional adjustments that reveal hidden maintenance costs in library evolution.

\subsection{Pre-trained Models and Upstream Perspectives}
Over the past decade, pre-trained models (PTMs) have become common in software systems, driven by the growing cost of training large models from scratch \cite{Geiping2023}. PTMs are trained in advance on vast datasets and made publicly available, enabling teams to achieve state-of-the-art results with fewer resources \cite{Pepe2024, Jiang2023}. Notable examples of PTMs are Phi-4 \cite{phi4}, Deepseek \cite{DeepSeek-AI2025}, and Stable Diffusion \cite{sd}. Registries such as Hugging Face, TensorFlow Hub, and PyTorch Hub have emerged as central hubs for open-source PTM dissemination, playing a similar role to the library managers that revolutionized software engineering \cite{Jiang2023}. The success of Hugging Face as a community PTM hub has been extensively documented by recent studies \cite{Jiang2023, Castano2024, Jones2024}. They lower the barrier to reuse while also introduce new questions about governance, versioning, and maintenance from the upstream perspective. 

Transparency and documentation remain major challenges in PTM ecosystems. Many PTMs lack clear documentation on training data, bias, licensing~\cite{Pepe2024}, ethics~\cite{Yang2024}, and versioning, making them difficult to identify, compare, and reuse reliably~\cite{Jiang2025, Ajibode2025}, while naming inconsistencies and poor metadata further complicate reuse~\cite{Taraghi2024} Several studies have identified attributes that facilitate PTM reuse. Drawing from developer interviews and ecosystem analysis, Jiang et al. \cite{Jiang2023} and Gong et al. \cite{Gong2023} identified several PTM attributes that enable effective reuse: provenance, reproducibility, portability, and clear licensing. Although ease of reuse has improved through the hubs, prior work highlighted concerns like PTM tampering, undocumented limitations, and weak enforcement of security practices \cite{Jiang2022, Montes2022, Castano2024}. Finally, environmental impact is another growing concern, yet only a small fraction of PTMs report training carbon emissions \cite{Castano2023}. Overall, despite the benefits of PTM hubs, substantial gaps remain in documentation quality, governance, and risk management.

\subsection{Pre-trained Model Reuse as Dependencies in Downstream Systems}

By leveraging PTMs, developers can achieve strong performance on new tasks without the cost of training models from scratch, reducing development time \cite{Han2021, Davis2023}. As PTMs proliferate, recent work characterizes them as a new class of software dependency whose learned behavior and integration requirements differ fundamentally from traditional libraries \cite{Yasmin2025}. This practice, referred to as \emph{PTM reuse}, involves integrating pre-trained models into downstream software systems \cite{banyongrakkul2025}. While powerful, PTMs also introduce fragility: their black-box nature can lead to software issues when downstream developers lack visibility into their behavior \cite{Taraghi2024}, and their deployment often requires substantial infrastructure and computational resources \cite{Zhou2023}. As a result, understanding how PTMs are reused, integrated, and maintained is increasingly critical to ensuring system maintainability and reliability.

Recent studies have begun to examine the challenges of PTM reuse from the downstream perspective. Davis et al. \cite{Davis2023} highlighted concerns around standardization and portability, while forum studies on StackOverflow revealed issues such as fairness concerns, parameter misconfigurations, version incompatibilities, and challenges in fine-tuning and interpreting PTM outputs \cite{Chakraborty2021, Tan2024}. Similarly, studies of the Hugging Face ecosystem reported issues with missing attributes, performance inconsistencies, and memory limitations \cite{Jiang2023, Taraghi2024}.

Recently, research has shifted towards downstream developers and their code bases. Pepe et al. \cite{Pepe2024} and Gao et al. \cite{Gao2025} identified transparency and safety gaps in PTM documentation and integration practices. Jiang et al. \cite{Jiang2024} provided large-scale metadata linking upstream PTMs to downstream OSS repositories, revealing documentation and licensing inconsistencies. Our prior work \cite{banyongrakkul2025} analyzed downstream OSS issue reports, identifying a diverse set of practical challenges, ranging from resource constraints to functionality enhancement, and found that PTM bugs persist longer than general software defects.

\subsection{Research Gaps and Motivations}

\textbf{PTMs as a new kind of software dependency}---PTMs are increasingly reused as long-lived dependencies rather than one-off assets, making their evolution an important software engineering concern for maintainability, reliability, and developer productivity. Unlike traditional software libraries, which are code-centric and relatively transparent, PTMs are data-driven components with opaque internals, learned behavior, and heavy resource demands. Recent work has conceptualized these differences as \emph{Software Dependencies~2.0}, highlighting that existing dependency management techniques offer limited guidance when the dependency is a trained model rather than a human-written library \cite{Yasmin2025}.

\textbf{The gap: from static snapshots to evolutionary dynamics}---While recent studies have explored PTM reuse from a downstream perspective, they primarily offer a static view of the ecosystem. They have successfully cataloged documentation gaps \cite{Pepe2024}, safety risks \cite{Gao2025}, and architectural patterns \cite{Yasmin2025} at specific points in time. However, these studies describe the challenges of reuse without capturing how reuse evolves as a project matures. Furthermore, it remains unknown whether PTM changes follow established patterns of traditional software maintenance or if the unique nature of PTMs necessitates an entirely different evolutionary trajectory.

\begin{figure}[htb]
    \centering
    \includegraphics[width=\linewidth]{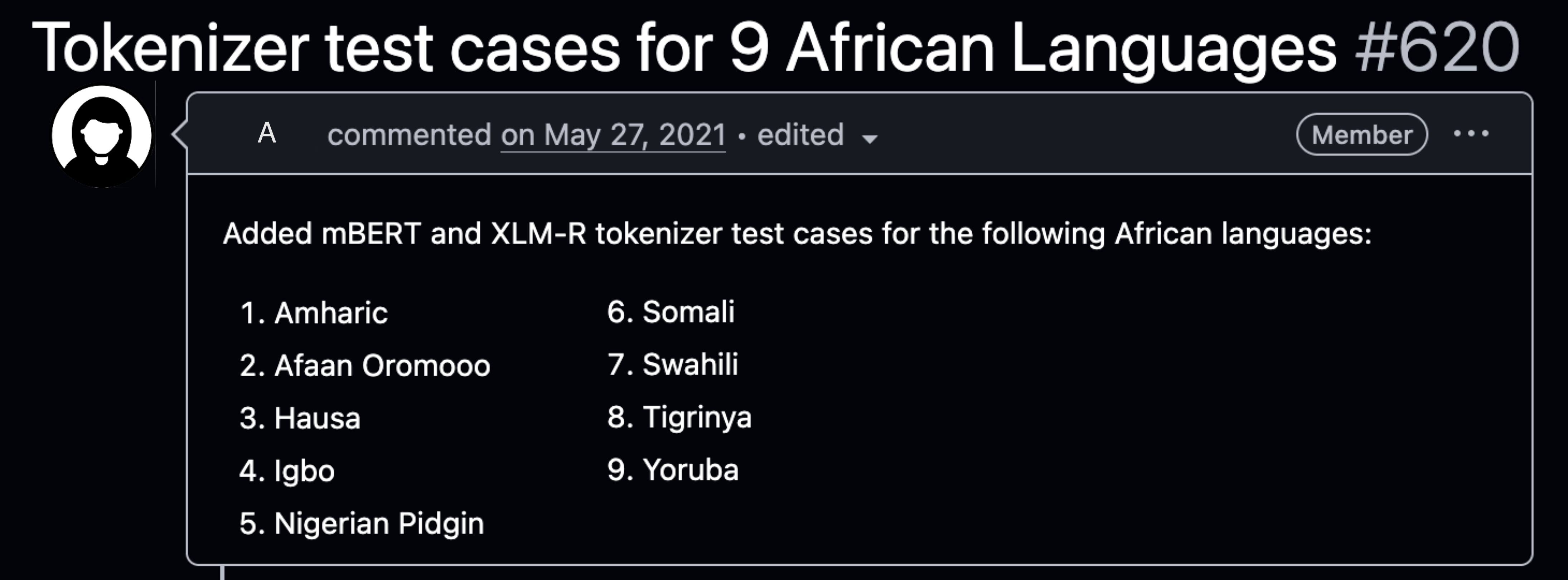}
    \caption{A motivating example where developers introduce new PTMs to test existing PTM-based tokenizer performance.}
    \label{fig:motivation}
\end{figure}

We currently lack empirical evidence regarding the longitudinal trajectory of PTMs in downstream OSS, including how often changes occur and what rationales drive them. As illustrated in Figure~\ref{fig:motivation}, developers sometimes introduce new PTMs (e.g., \textit{bert-base-multilingual-uncased}) specifically to validate existing PTM-based tokenizer behaviors for newly supported languages. Yet, it is unclear whether this reflects a systematic pattern, an isolated workaround, or one of many diverse rationales shaping PTM evolution across projects. While traditional libraries are often updated for bug fixing and compatibility, this example suggests that PTM evolution may be motivated by different drivers (e.g., the inherent challenges of behavior validation). Without understanding how often such changes occur and what rationales developers document to justify them, we cannot anticipate the long-term maintenance needs (e.g., technical debt accumulation) of downstream systems that depend on PTMs.

\textbf{Methodological limitations}---Addressing this gap also requires overcoming methodological limitations in existing datasets and approaches. To the best of our knowledge, while large-scale studies (e.g., \cite{Pepe2024, Jiang2024}) provide valuable foundations for identifying PTM reuse, they are not designed particularly for longitudinal analysis and require improvement in coverage and quality. Furthermore, many PTMs act as multi-role components \cite{Yasmin2025} and do not offer clear one-to-one replacement relationships \cite{Ajibode2025}. As a result, traditional dependency comparison techniques (e.g., \cite{Teyton2014, He2021, Islam2023}) are insufficient for detecting PTM changes across software versions, limiting our ability to study PTM evolution at scale.

\textbf{Research objective}---These gaps motivate a comprehensive investigation of PTM reuse focused on continuous change in downstream projects. Our objective is to quantify PTM changes in direct comparison to traditional libraries, examining how developers document and justify these transitions. By identifying change patterns and the unique rationales behind these evolutions, we aim to provide a comprehensive understanding of how model-centric dependencies age within real-world OSS systems and seek to identify areas where current software engineering practices and tools may be insufficient for supporting the PTM-integrated OSS lifecycle.

\section{Research Questions}
\label{sect:rqs}
To address these gaps, we formulate the following research questions. 

\textbf{RQ1: How do PTM-related changes occur in downstream projects, and how do they compare to traditional software library changes?}

This question quantifies how often PTMs change, what evolution patterns emerge, and how these compare to libraries.

\noindent{\small $\bullet$} \textbf{RQ1.1:} \textit{How frequently do PTM usages change, and how is the distribution of change types compared to libraries?} This sub-question examines how often PTM usage changes occur in downstream OSS projects, their cadence relative to traditional libraries, and how their distribution across additions, removals, and migrations compares with library changes. We answer it by comparing consecutive software versions and extracts PTM snapshots to identify changes between versions.

\noindent{\small $\bullet$} \textbf{RQ1.2:} \textit{How does PTM reuse evolve over time compared to library reuse?} We aim to understand the temporal evolution of PTM reuse in downstream projects and to compare it with the evolution of traditional library reuse. This examines whether PTM adoption and reuse exhibit distinct growth patterns over time. We analyze longitudinal trends from first adoption to subsequent growth and compare them with library reuse.

\textbf{RQ2: To what extent are PTM-related changes documented, and what rationales are evident?}

Building on RQ1, this question examines how developers communicate and justify PTM changes, and how this compares to library changes.

\noindent{\small $\bullet$} \textbf{RQ2.1:} \textit{How frequently are PTM changes documented, and where does documentation appear?} This sub-question examines how often PTM changes are documented in downstream projects and where such documentation appears in the development workflow. We answer it by manually analyzing textual artifacts associated with detected PTM changes, such as release notes, commit messages, and pull request descriptions.

\noindent{\small $\bullet$} \textbf{RQ2.2:} \textit{What rationales do developers document to justify PTM changes?} We examine the rationales that developers provide when PTM changes are documented. We answer it through a qualitative analysis of the collected textual artifacts to identify and organize the reasons that drive PTM additions, removals, and migrations.

\noindent{\small $\bullet$} \textbf{RQ2.3:} \textit{How do documentation practices and reported rationales differ between PTM and library changes?} This compares PTM and library changes in terms of documentation coverage and reported rationales. It reveals how documentation practices vary and highlights the motivations unique to PTM changes compared to library changes.

\section{Dataset}
\label{sect:dataset}
\begin{figure*}
    \centering
    \includegraphics[width=\linewidth]{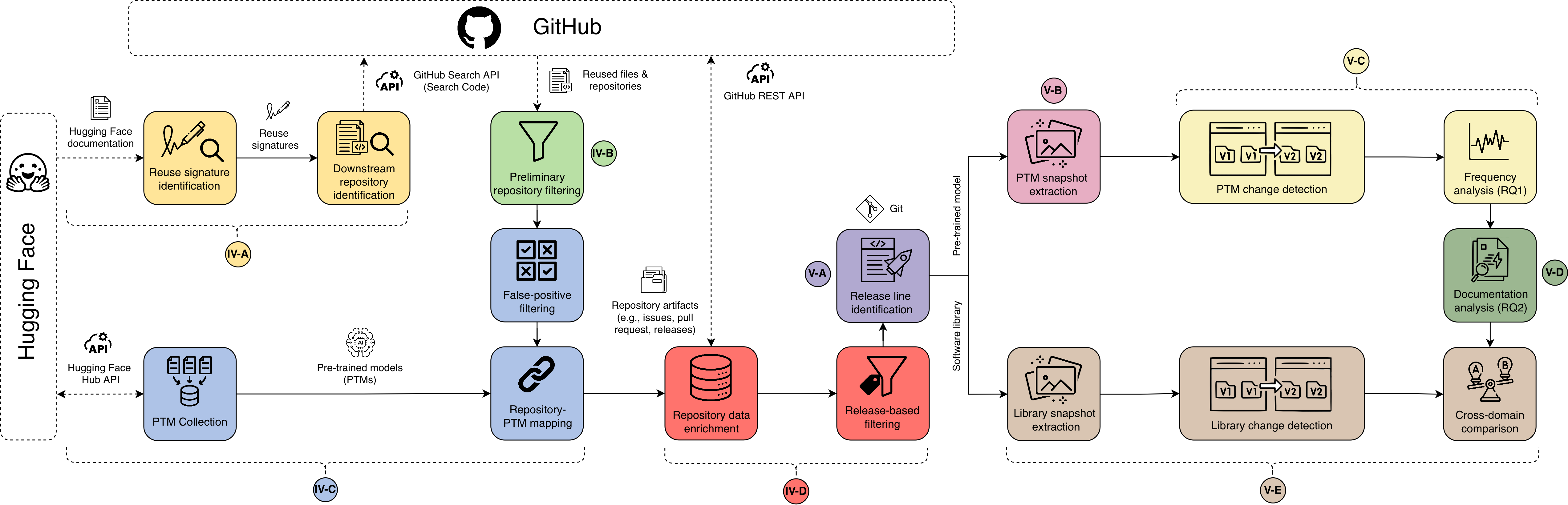}
    \caption{Overview of our dataset creation and analysis pipeline. The process spans from initial PTM signature identification (Section \ref{sect:dataset}) to the heuristic detection and cross-domain comparison of PTM and library changes (Section \ref{sect:method}). Labels (e.g., IV-A, V-B) correspond to specific subsections detailing the methodology.}
    \label{fig:approach}
\end{figure*}

This section describes the construction of our dataset used to analyze PTM changes in downstream projects. Figure~\ref{fig:approach} provides an overview of both the dataset construction and methodology. Our dataset construction consists of four phases: (1) downstream repository identification, (2) preliminary filtering, (3) static validation and repository–PTM mapping, and (4) data enrichment and final filtering. Detailed procedures and additional examples are provided in the online appendix\textsuperscript{\ref{fn:replication}}.

\subsection{Downstream Repository Identification}
To identify GitHub OSS projects reusing open-source PTMs, we defined a set of \emph{reuse signatures}, i.e., code patterns indicating PTM loading or instantiation. Following \cite{Jiang2024}, each signature consists of (1) a library import statement (e.g., \texttt{transformers}) and (2) a function or method call that loads a PTM (e.g., \texttt{from\_pretrained()}). To ensure coverage and currency, we manually reviewed the official Hugging Face documentation\footnote{\url{https://huggingface.co/docs}}, which catalogs all libraries supporting Hugging Face PTM reuse. The first author and another author independently reviewed the documentation, with each library's API Reference section examined to identify all function calls related to model loading. The extracted signatures were then combined and discussed to resolve differences. The extraction procedure was discussed and validated by all authors. This process yielded 355 signatures across 26 Python libraries.

We then conducted large-scale searches across public GitHub OSS repositories using the GitHub Search API (Search Code endpoint\footnote{\url{https://docs.github.com/en/rest/search/search\#search-code}}) to identify files containing any reuse signature, which we refer to as \emph{reused files}. Compared with alternatives like Sourcegraph used in \cite{Jiang2024} and GitHub Dependency Graph used in \cite{Pepe2024}, the GitHub Search API provides broader coverage and direct programmatic access through Python SDKs. Additionally, to improve coverage, we executed each search query using both import-style and from-style variations of the signatures. Using this approach, we identified 730,891 reused files across 239,586 candidate downstream repositories potentially reusing PTMs between June and July 2025.

\subsection{Preliminary Repository Filtering}
To facilitate subsequent steps and reduce unnecessary processing overhead, we performed a preliminary filtering step to retain repositories of acceptable quality. Following common GitHub-based quality indicators \cite{Toma2024, Gonzalez2020} also used in our prior work on PTM-based downstream software challenges~\cite{banyongrakkul2025}, we applied several criteria, including data availability, non-zero repository size, popularity (at least five stargazers or five forks), and main programming language (Python). Python is the most popular language for AI-related projects \cite{Gonzalez2020}. We also assessed project activeness based on recent maintenance activity (last push date on or after January 1, 2025) and performed keyword-based filtering on repository titles, topics, and descriptions to exclude educational materials, tutorials, and homework submissions. With this process, we ended up with 10,260 repositories with 56,462 reused files.

\subsection{Static Validation and Repository-PTM Mapping}
\label{subsect:static_analysis}

In this step, we performed \emph{static analysis} on the identified reused files to validate PTM reuse and establish repository–PTM mappings. Because the GitHub Search Code API relies on string matching, additional validation was needed to reduce false positives. To assess this, we manually inspected 383 reused files, sampled with a 95\% confidence level and a 5\% margin of error. The first author conducted the manual inspection, and representative cases of false positives and PTM–repository mappings (i.e., how PTM identifiers are passed through method calls) were discussed with all authors to validate the categorization. Among these, we identified 73 false positives (19\%), which fall into four categories: (1) signatures inside comments, (2) similarly named methods from unrelated packages (e.g., matching \texttt{load()} from \texttt{json} instead of \texttt{spacy}), (3) PTM loading in example or demonstration files, and (4) files from third-party packages (e.g., \texttt{site-packages/}) rather than project-authored code.

Our static analysis extended the pipeline proposed by \cite{Jiang2024}, which uses the Scalpel framework\footnote{\url{https://python-scalpel.readthedocs.io/en/latest/\#}} to parse source code into an Abstract Syntax Tree (AST) and extract imports and function calls. While the original approach mainly relied on text-based matching (i.e., PTM identifiers provided as direct string literals in function calls) and addressed only the first two false-positive cases, our version introduced variable resolution to track PTM identifiers stored in variables, along with additional filtering rules to handle all four categories identified in the manual analysis. We evaluated these improvements using the same sampled set of 383 files described above. Compared to the baseline approach \cite{Jiang2024}, our method reduced false positives by 52.08\% and increased reused-file coverage by 64.00\%.

We collected open-source PTMs from the Hugging Face PTM hub, the largest public registry of PTMs \cite{Jiang2023}. Using the Hugging Face Hub API\footnote{\url{https://huggingface.co/docs/hub/en/api}}, we extracted PTM identifiers, names, and metadata to construct a searchable index. In total, we identified 1,805,282 PTMs from Hugging Face as of June 2025, representing an increase of 541.82\% over PeaTMOSS \cite{Jiang2024} (August 2023). To enable more accurate repository--PTM mapping, we redesigned the static analysis pipeline with several enhancements. For example, we enabled tracking of PTM loading when their names were stored in variables while preserving assignment order. We also added sub-module--aware import resolution (e.g., \texttt{from transformers.models.auto import AutoModel}). We further supported conditional reassignments (e.g., \texttt{if--else} blocks), since projects may select among interchangeable PTMs depending on runtime conditions (e.g., based on PTM size or resource constraints)~\cite{Yasmin2025}. Finally, we extended variable resolution to capture calls made through object attributes or class defaults. After applying static analysis, we identified 4,067 downstream repositories reusing 3,018 PTMs across 15,599 reused files.

\subsection{Data Enrichment and Final Filtering on Releases}
\label{subsect:final_filtering}
For each validated repository, we collected project artifacts such as commit histories, releases, pull requests, and issue reports, stored in our local MySQL database. To ensure that the final dataset reflected mature and systematic software engineering practices, particularly in release management \cite{Laukkanen2018}, we applied five release-based exclusion criteria adapted from previous work on GitHub releases \cite{Coelho2018, Joshi2019, Kilic2023, Chakroborti2023}.

\begin{figure}[!ht]
    \centering
    \includegraphics[width=\linewidth]{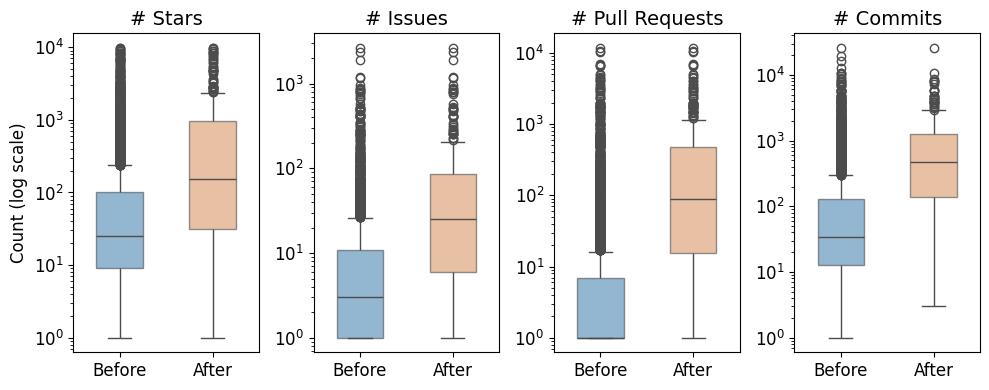}
    \caption{Distribution of repository characteristics before and after the release-based filtering in Section~\ref{subsect:final_filtering} (log scale).}
    \label{fig:release_filtering}
\end{figure}

After removing repositories without any releases, we first excluded draft and prerelease tags as they represent unstable or incomplete builds. Second, repositories with fewer than two releases were removed since at least two releases are required to observe PTM changes. Third, to filter out irregular release practices \cite{Kilic2023}, we excluded repositories whose median release interval was outside 7–365 days, as well as those with extremely high  or very low release activity. Fourth, repositories without releases in the most recent year (after January 1, 2025) were excluded as inactive. Finally, we filtered repositories based on semantic versioning (SemVer) compliance \cite{SemVer} using regular expressions, as inconsistent tag naming may indicate poor release practices (see details\textsuperscript{\ref{fn:replication}}). After applying all criteria, the dataset comprised 341 repositories with 832 reused files and 6,242 releases, which were retained for our analyses. To characterize the nature of this filtered dataset, we statistically compared it against the broader pool of 4,067 confirmed PTM-reusing repositories using four measures: stargazers, issues, pull requests, and commits. As shown in Figure~\ref{fig:release_filtering}, Mann--Whitney U tests \cite{Mann1947} revealed significant differences across all measures ($p < 0.001$), with medium effect sizes measured using Cohen’s $d$ (stargazers: -0.62; commits: -0.49; pull requests: -0.38; issues: -0.37) \cite{Diener2010}, confirming that our criteria inherently favor more mature and active projects, which is an intentional design choice, as longitudinal analysis requires reliable release histories.

\section{Methodologies}
\label{sect:method}
This section describes the methodological framework used to analyze PTM changes and compare them with library changes in downstream GitHub OSS repositories. We used tagged versions (i.e., releases) as the unit of analysis, as they represent stable software versions delivered to users and are commonly used in prior studies~\cite{Teyton2014, Terzi2022, Bavota2015}. Building on the dataset introduced in Section~\ref{sect:dataset}, our approach comprises five main components: (1) identifying development lines from repository histories, (2) extracting PTM snapshots from releases, (3) detecting PTM changes between consecutive releases, (4) analyzing PTM change documentation, and (5) constructing a library change baseline for cross-domain comparison, as illustrated in Figure~\ref{fig:approach}.

\subsection{Release Line Identification}
OSS repositories often maintain multiple concurrent release branches, for example, release v1.x.y and v2.x.y developed in parallel. Each branch forms its own evolution trajectory. To avoid incorrect ordering of PTM changes based only on release dates, we identified \emph{release lines} (i.e., ordered sequences of releases belonging to the same development lineage). We first retrieved all repository branches, resolving each release tag to its corresponding commit SHA. For each branch-tag pair, we checked whether the tag commit was an ancestor of the branch tip (using \texttt{git merge-base --is-ancestor}). These candidate branches (i.e., shortlisted branches) linked to releases and filtered out branches unrelated to any tagged version.

Since branch reachability alone may be affected by merge commits, we performed \emph{first-parent traversal} on each candidate branch (using \texttt{git rev-list --first-parent}), producing a linear history representing the main development path while ignoring side branches merged later. A release tag was assigned to a branch if its commit appears in this sequence, with its position determining the release order within the line. We excluded repositories with incomplete or ambiguous coverage (e.g., tags unreachable from any shortlisted branch) and release lines with fewer than two releases. From this process, we identified 557 release lines with 4,988 releases across 323 downstream repositories, which are included as a MySQL database in our replication package\textsuperscript{\ref{fn:replication}}.

\subsection{PTM Snapshot Extraction}
The second stage of our methodology extracted PTM snapshots for every tagged release to capture the exact state of PTM reuse at each release point. For each reused file in a release, we applied the static analysis that we used in the phase of data validation and repository-PTM mapping, described in Section \ref{subsect:static_analysis}. The process parsed the source code into an AST and extracts import statements that matched known signatures. It then applied variable-assignment tracing and function-call detection to resolve PTM identifiers accurately. We also accounted for file renaming using commit history information. The output of this file-level analysis was a snapshot of all detected PTM occurrences within that file version, which were subsequently aggregated to the release level for change detection.

\subsection{PTM Change Detection}
\label{subsect:change_detection}

This stage identified PTM change events by comparing PTM snapshots across consecutive releases within each release line. To support precise reasoning, we formalized the release-level PTM snapshot as a \emph{multiset} of PTM identifiers. Let $\mathcal{M}$ be the universe of all PTM identifiers detected in the repository. For a given release $R$, we define the snapshot $S_R$ as the multiset sum of PTMs found across all reused files $f \in \text{Files}(R)$:
\[
S_R = \biguplus_{f \in \text{Files}(R)} S_{f,R}
\]

where $\uplus$ denotes a multiset union that preserves counts. Consequently, the total occurrences of a specific PTM $m \in \mathcal{M}$ in release $R$ is given by $\text{count}_R(m) = \sum_{f} \text{count}_{f,R}(m)$. This multiset approach is necessary because the same PTM may be reused for different tasks or roles within a project; treating them as individual instances ensures that changes to one usage are not obscured by the stability of another. Using these snapshots, we traversed each release line in chronological order. For every pair of adjacent releases $(R_i, R_{i+1})$ (i.e., a \emph{release pair}), we computed the evolution of PTM usage by defining the multiset delta for each PTM $m$, $\Delta(m) = \text{count}_{R_{i+1}}(m) - \text{count}_{R_i}(m)$. Based on these deltas, we calculated the total number of added ($A$) and removed ($R$) PTM instances across the release pair as:
\[
A = \sum_{m \in \mathcal{M}} \max(\Delta(m), 0), \qquad R = \sum_{m \in \mathcal{M}} \max(-\Delta(m), 0)
\]
We categorized these changes into three distinct types:

\noindent{\small $\bullet$} \textbf{Additions:} Occur when a PTM's count increases ($\Delta(m) > 0$), covering both initial adoption and the introduction of existing PTMs to new call sites.

\noindent{\small $\bullet$} \textbf{Removals:} Occur when a PTM's count decreases ($\Delta(m) < 0$), representing either a total deletion or a reduction in the number of call sites.

\noindent{\small $\bullet$} \textbf{Migrations:} Following common manual validation practices \cite{He2021, Teyton2014, Islam2023}, we first established a baseline for potential migrations as $U = \min(A, R)$. We then identified \emph{migration candidates} by requiring that an addition and removal occur within the same file and commit. Finally, we performed manual validation to confirm the developer's intent; a candidate was only classified as a migration (annotated as ``Y'') if the code changes or commit messages explicitly indicated it (e.g., ``migrate from \textit{<model\_A>} to \textit{<model\_B>}''). Otherwise, they were recorded as independent additions and removals.

For each release pair, we recorded both scalar summaries and structured metadata, such as changed PTM identifiers, file-level locations, and first adoption flags, to support both quantitative and qualitative analyses. To ensure that only meaningful release pairs were included, we excluded transitions where neither $R_i$ nor $R_{i+1}$ contains any PTMs, retaining only pairs from the point of first PTM adoption onward. We refer to this point as $t_1$. After this filtering, our dataset comprised 3,265 releases across 451 release lines from 310 downstream repositories, yielding 2,814 valid release-pair comparisons.

\begin{figure}
    \centering
    \includegraphics[width=0.9\linewidth]{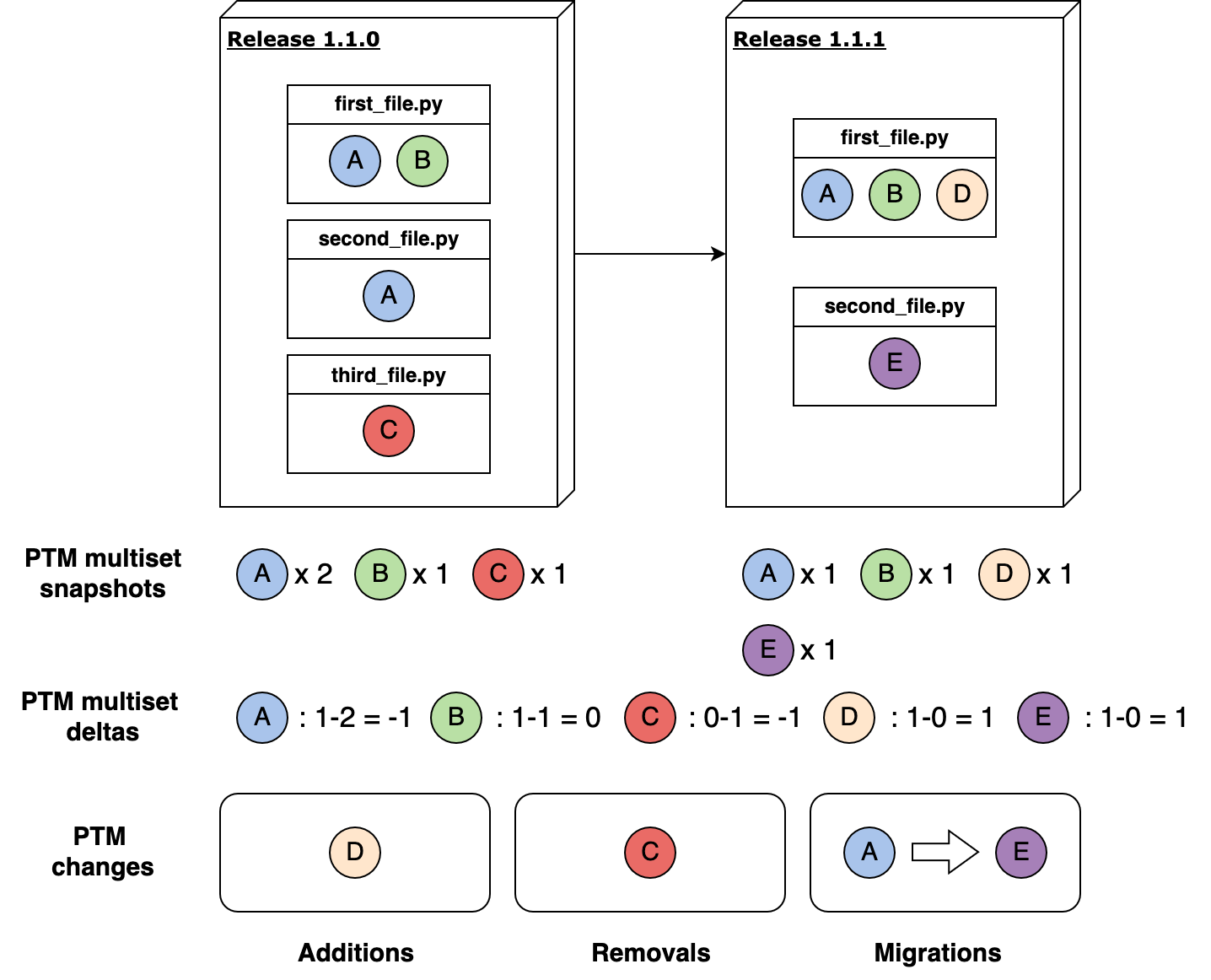}
    \caption{Example of PTM change detection using multiset snapshots across two consecutive releases (i.e., a \emph{release pair}), illustrating additions, removals, and a migration.}
    \label{fig:change_example}
\end{figure}

\noindent\textbf{Example.}
Figure~\ref{fig:change_example} illustrates a PTM change event across a release pair $(R_i, R_{i+1})$. In release $R_i$ (version 1.1.0), the snapshot is $S_{R_i} = \{\textit{A}:2,\; \textit{B}:1,\; \textit{C}:1\}$, where PTM \textit{A} appears twice, while \textit{B}
and \textit{C} each appear once. In the subsequent release $R_{i+1}$ (version 1.1.1), the snapshot becomes $S_{R_{i+1}} = \{\textit{A}:1,\; \textit{B}:1,\; \textit{D}:1,\; \textit{E}:1\}$.
Comparing these snapshots, we obtain the multiset deltas for all PTMs:
\[
\begin{gathered}
\Delta(\textit{A})=-1,\quad
\Delta(\textit{B})=0,\quad
\Delta(\textit{C})=-1,\\
\Delta(\textit{D})=+1,\quad
\Delta(\textit{E})=+1
\end{gathered}
\]

The total number of additions and removals are $A = 2$ and $R = 2$, yielding a baseline of $U = \min(A, R) = 2$ migration candidates.
Concretely, the totals are computed as
\[
A = \max(1,0) + \max(1,0) = 2 \; (\textit{D}, \textit{E}),
\]
\[
R = \max(1,0) + \max(1,0) = 2 \; (\textit{A}, \textit{C}).
\]

One candidate corresponds to a substitution from \textit{A} to \textit{E} within the same file. The remaining changes---addition of \textit{D} and removal of \textit{C}---do not meet the migration criteria and are classified as independent addition and removal events.

\subsection{Qualitative Analysis on PTM Change Documentation}
We conducted a qualitative analysis to examine how downstream developers document PTM changes, focusing on whether changes are explicitly documented where documentation appears, and what motivations developers state. The analysis consisted of two phases: (1) textual artifact collection and (2) qualitative coding.

\subsubsection{\textbf{Relevant Textual Artifact Gathering}}
For each release pair with detected PTM changes, we collected related documentation, including release notes, commit messages, pull request descriptions, issue discussions, code comments, and Markdown-based project files. For commit messages, we restricted collection to commits that modified files where PTM changes were detected, replaying commits chronologically and retaining only those affecting PTM usage. Pull requests and issues were included when linked from releases or commits. Additional artifact types were identified through a pilot analysis, which also revealed that PTM changes may appear in Markdown files (e.g., \texttt{README.md}, \texttt{CHANGELOG.md}) and code comments near modified lines.

\subsubsection{\textbf{Qualitative Analysis Procedure}}
After collecting the textual artifacts, we manually annotated them in two steps. First, we labeled whether it explicitly documents a PTM change (``with documentation'' or ``without documentation''). Second, for documented changes, we labeled whether a motivation was stated (``with rationale'' or ``without rationale''). To align annotation criteria, the first author randomly sampled 50 artifacts and conducted independent annotation with another author. Inter-rater agreement scores were 0.87 for documentation identification and 0.76 for rationale identification, indicating substantial agreement \cite{Landis1977, McHugh2012}. Disagreements were discussed to refine the annotation guidelines, which were then applied by the first author to annotate the remaining artifacts. 

\begin{lstlisting}[
  style=no_rationale,
  caption={Examples of PTM change documentation without explicit rationale (sources: \cite{commit_roberta_removal, commit_swin_addition}).},
  label={lst:no_rationale}
]
(*\underline{\textbf{[Example 1]}}*)
(*\textbf{Artifact type}*) : Commit message
(*\textbf{Changed PTM}*)   : FacebookAI/roberta-base
(*\textbf{Change type}*)   : Removal
(*\textbf{Raw text}*)      : "Remove superseeded roberta.py script"
(*\textbf{Changed code}*)  :
(*\color{red}{- model\_name = "FacebookAI/roberta-base"}*)
(*\color{red}{- hf\_model = AutoModelForMaskedLM.from\_pretrained(model\_name)}*)
(*\underline{\textbf{[Example 2]}}*)
(*\textbf{Artifact type}*) : Commit message
(*\textbf{Changed PTM}*)   : openmmlab/upernet-swin-large
(*\textbf{Change type}*)   : Addition
(*\textbf{Raw text}*)      : "Swin Upernet model added"
(*\textbf{Changed code}*)  : 
(*\color{green!60!black}{+ self.swin\_upernet\_model = UperNetForSemanticSegmentation
\\.from\_pretrained("openmmlab/upernet-swin-large")}*)
\end{lstlisting}

In the final annotation procedure, the first author followed a structured decision process: (1) search for PTM-related keywords, including corresponding PTM identifiers and change-related terms (e.g., add, remove, migrate); (2) if no match is found, manually scan the artifact content for up to three minutes; (3) if neither step yields evidence, annotate the artifact as ``without documentation''; and (4) if documentation is identified, carefully review the relevant content and annotate it as ``with rationale'' if an explicit motivation or purpose for the PTM change is stated. For artifacts annotated as ``with rationale'', we first extracted key information describing the stated rationale from the raw text, using the surrounding code changes as contextual support. Note that a single textual artifact may contain zero, one, or multiple rationales, and a single rationale may involve multiple PTM change events. In contrast, artifacts annotated as ``without rationale'' typically consist of only a brief action verb and model name with no further explanation, as shown in Listing~\ref{lst:no_rationale}.

\begin{figure}[htb]
    \centering
    \includegraphics[width=\linewidth]{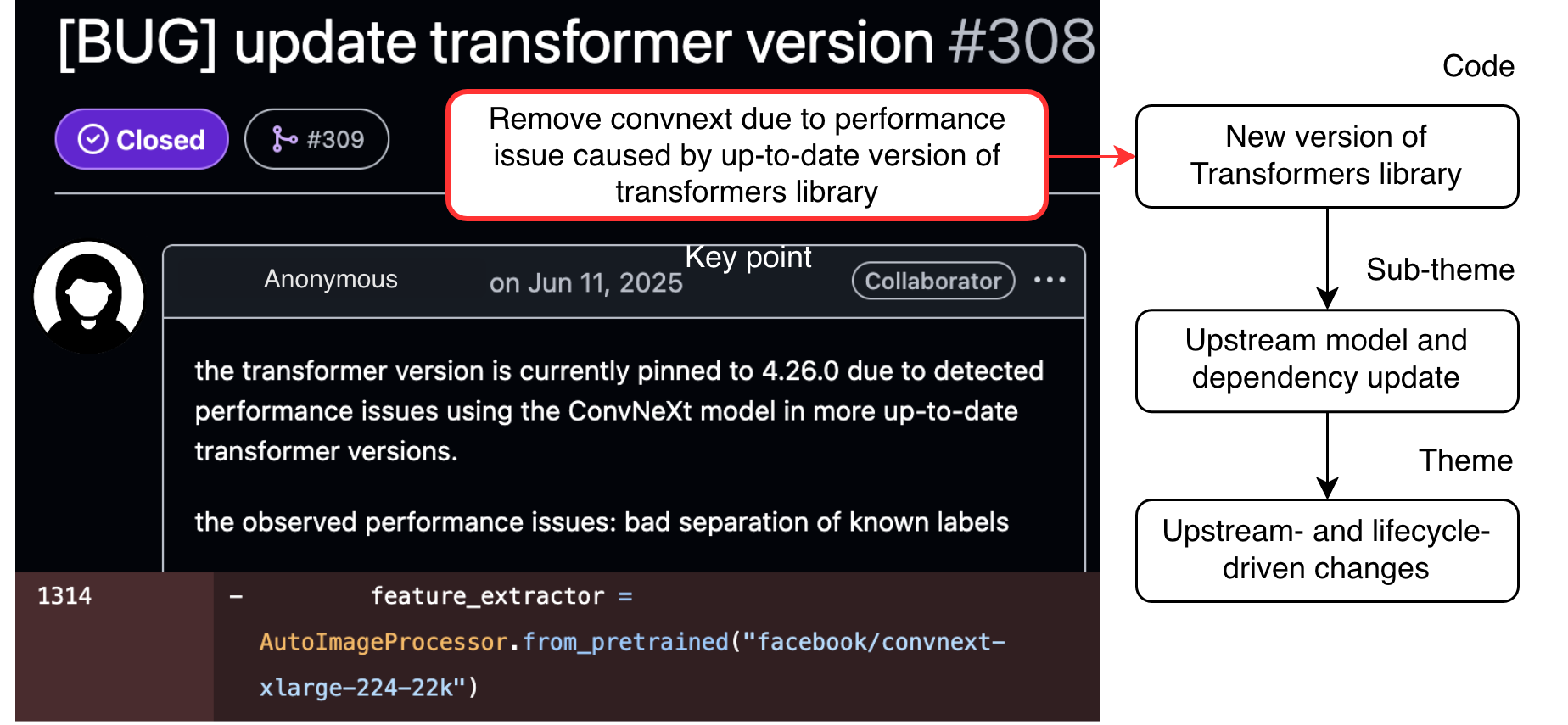}
    \caption{Example of the qualitative process used to analyze documented rationales of PTM changes.}
    \label{fig:trigger_example}
\end{figure}

We then applied \emph{inductive open coding}~\cite{thomas2006} to group similar rationales into categories and assign descriptive labels (i.e., \emph{codes}). All analyzes were conducted using shared spreadsheets\textsuperscript{\ref{fn:replication}} accessible to all authors. We followed a bottom-up approach, iteratively comparing newly identified rationales with existing codes and refining the coding scheme as new patterns emerged (see Figure~\ref{fig:trigger_example}). We began by assigning an initial set of low-level codes to the extracted rationale information from a sample of 30 rationales. To maintain quality and consistency, the coding then proceeded in batches of 50 rationales, reusing existing codes whenever possible, and introducing new codes only when no suitable code existed. Related codes were grouped into higher-level categories (i.e., sub-themes and themes) to capture broader patterns~\cite{banyongrakkul2025}. The coding scheme was continuously refined through regular discussion and consensus among all authors. The process was closely monitored in weekly meetings, during which codes were adjusted, merged, split, or removed as needed. Any disagreements or ambiguities were resolved through discussion until consensus was reached.

\begin{table*}[!htb]
\centering
\scriptsize
\caption{Definitions of documentation and rationale coverage metrics for PTM and library changes. All values are in $[0, 1]$.}
\label{tab:doc_metrics}
\renewcommand{\arraystretch}{0.9}
\begin{tabular}{clll}
\toprule
\textbf{ID} & \textbf{Metric} & \textbf{Definition} & \textbf{Calculation} \\
\midrule
1 & Documentation rate & Proportion of change events that are documented & \# documented changes / \# total changes \\
2 & Release-pair documentation rate & Proportion of release pairs with documented changes & \# documented release pairs / \# release pairs with changes \\
3 & Rationale rate & Proportion of change events with documented rationales & \# changes with documented rationales / \# total changes \\
4 & Rationale rate of documented changes & Proportion of documented changes that include rationales & \# changes with documented rationales / \# documented changes \\
5 & Rationale rate of artifacts & Proportion of document artifacts that contain rationales & \# artifacts with documented rationales / \# total artifacts \\
\bottomrule
\end{tabular}
\end{table*}

\subsection{Library Change Baseline Construction \& Comparison}
To enable cross-domain comparison throughout RQ1 and RQ2, we mirrored our PTM release-line pipeline on traditional libraries within the same dataset, temporally anchored at $t_1$ (first PTM adoption) to ensure that both domains are tracked over the same release window. Across the identical set of 2,814 valid release pairs, we followed Islam et al.~\cite{Islam2023} to detect library changes from four Python dependency specification files: \texttt{requirements.txt}, \texttt{environment.yml}, \texttt{Pipfile}, and \texttt{pyproject.toml}. We extended their approach by categorizing changes into additions, removals, version updates, and migrations. Migration candidates were identified when an addition and removal co-occured within the same file and commit, then validated in two steps: (1) matching against a curated list of analogous library pairs by prior work~\cite{Islam2023, Islam2024}, and (2) applying a GPT-based validation prompt~\cite{Islam2024} for the remaining candidates. Invalid migration candidates were reclassified as separate addition and removal events.

To compare PTM and library change patterns in RQ1, we applied the same analyses to both domains in parallel. For RQ1.1, we compared change frequency (proportion of release pairs with at least one change), type distribution, and per-release-line cadence (releases per change event) between PTM and library changes. To statistically confirmed cadence differences, we applied two complementary non-parametric tests with Bonferroni correction ($\alpha = 0.05/4 = 0.0125$): (1) paired Wilcoxon signed-rank tests~\cite{Woolson2008} on release lines exhibiting at least one change of each type in both domains, and (2) Mann-Whitney U tests~\cite{Mann1947} at the population level. Effect sizes are reported as Rank-Biserial $r$ and Cliff's $\delta$, respectively. For RQ1.2, we compared the temporal evolution by applying the same post-adoption growth analysis to library release lines. We computed the median growth factors and addition-only line rates for both domains. To analyze when additions concentrate across the release-line lifetime, we applied a six-stage lifecycle normalization to both PTM and library addition events, dividing each release line into equal stages (0--16\%, 16--33\%, 33--50\%, 50--66\%, 66--83\%, 83--100\%) from first to latest release, retaining only lines with at least six releases, and counting addition events within each stage.

To compare documentation and rationale practices between PTM and library changes in RQ2, we performed a lightweight but representative manual annotation on a stratified random sample of library change events. To achieve 95\% confidence with a $\pm$5\% margin of error, a minimum sample of 374 events is required. We increased this to 420 events to ensure stable stratified analysis, distributed as follows: 140 additions, 140 updates, 90 removals, and 50 migrations. For each sampled event, we applied the same labeling protocol used for PTM changes: identifying whether the change was accompanied by documentation and whether an explicit rationale was present across the same set of associated textual artifacts. Documentation and rationale rates derived from this sample were then compared directly against the PTM results. To systematically compare both domains, we defined five metrics that captured documentation and rationale coverage at different granularities, summarized in Table~\ref{tab:doc_metrics}.

We then followed a qualitative approach similar to the one used for the PTM rationales. We carefully read each artifact, extract key points, and assign low-level codes to capture the underlying rationale. Since the rationales behind library changes have been extensively studied in prior literature~\cite{Teyton2014, Kabinna2016, He2021, Barbosa2022, Jaisri2025, Bavota2015}, we reused and adapted existing sub-themes and themes rather than deriving them entirely from scratch. We selected prior studies that systematically derive and clearly report library change taxonomies, ensuring that the categories are well-defined and empirically grounded. Specifically, we examined sub-themes and themes from prior work based on their descriptions and examples, aligning our codes accordingly. Each may map to none, one, or multiple codes depending on conceptual overlap. The resulting codes were then grouped into high-level themes using terminology consistent with prior work, forming the library rationale taxonomy for cross-domain comparison in RQ2.3.

\section{Frequency of PTM Changes (RQ1)}
\label{sect:result1}
To answer RQ1, we analyzed PTM changes at release-pair granularity and compare these observed patterns with library changes. As described in Section~\ref{subsect:change_detection}, we focused our analysis on the AI-active phase of each project's lifecycle, beginning at $t_1$. This resulted in 2,814 valid release-pair comparisons across 451 release lines from 310 repositories. By anchoring both domains to this shared observation window, we ensured that the subsequent comparisons of frequency and cadence were not biased by the longer historical presence of traditional libraries. Table~\ref{tab:dep_baseline} summarizes the PTM and library results used for cross-domain comparison throughout RQ1 and RQ2; detailed discussion follows in the respective subsections.

\begin{table}[!t]
\centering
\scriptsize
\caption{Cross-domain comparison summary of PTM and library change patterns across RQ1 and RQ2.}
\label{tab:dep_baseline}
\renewcommand{\arraystretch}{1.0}
\begin{tabular}{lrr}
\toprule
\textbf{Metric} & \textbf{PTM} & \textbf{Library} \\
\midrule
\multicolumn{3}{l}{\textit{Dataset overview --- Shared observation window ($t_1$ onwards)}} \\
\midrule
Total repositories                & \multicolumn{2}{c}{310} \\
Total release lines               & \multicolumn{2}{c}{451} \\
Total releases analyzed           & \multicolumn{2}{c}{3,265} \\
Total release-pair comparisons    & \multicolumn{2}{c}{2,814 (3,265-451)} \\
\midrule
\multicolumn{3}{l}{\textit{RQ1.1 --- Change frequency \& type distribution (Section~\ref{subsect:rq1.1})}} \\
\midrule
Release pairs with changes        & 14.43\% (406/2,814)  & 38.13\% (1,073/2,814) \\
Total change events           & 1,426                & 14,413 \\
Addition share                    & 88.57\% (1,263)     & 37.65\% (5,426) \\
Removal share                     & 8.34\% (119)        & 23.12\% (3,334) \\
Migration share                   & 3.09\% (44)         & 0.35\% (50) \\
Version update share                      & ---                 & 38.88\% (5,608) \\
\makecell[l]{Change rate\\(Release lines, $\geq$1 change)} & 48.1\%  (217/451)           & 56.3\% (254/451) \\
\makecell[l]{Addition rate\\(Release lines, $\geq$1 addition)} & 44.35\%  (200/451)           & 49.45\% (223/451) \\
\makecell[l]{Removal rate\\(Release lines, $\geq$1 removal)} & 8.20\%  (37/451)           & 35.48\% (160/451) \\
\makecell[l]{Migration rate\\(Release lines, $\geq$1 migration)} & 5.10\%  (23/451)           & 7.32\% (33/451) \\
Median addition cadence           & 3.75 releases/event     & 2.00 releases/event \\
Median removal cadence            & 7.00 releases/event     & 3.50 releases/event \\
Median migration cadence          & 6.50 releases/event     & 8.00 releases/event \\
\midrule
\multicolumn{3}{l}{\textit{RQ1.2 --- Temporal evolution of reuse (Section~\ref{subsect:rq1.2})}} \\
\midrule
Release lines with net growth     & 39.2\% (85/217)              & 58.3\% (148/254) \\
Addition-only growth lines\textsuperscript{$\diamond$}        & 41.2\% (35/85)           & 12.2\% (18/148) \\
Median growth factor              & 2.0$\times$         & 1.29$\times$ \\
\midrule
\multicolumn{3}{l}{\textit{RQ2 --- Documentation \& documented rationales of PTM changes (Section~\ref{sect:result2})}} \\
\midrule
Total changes analyzed             & 1,426        & 420\textsuperscript{$\diamond\diamond$} \\
Total artifacts identified        & 417                  & 485 \\
Documentation rate & 37.94\% (541/1,426)            & 66.67\% (280/420)\\
\makecell[l]{Release-pair\\documentation rate}   & 40.64\% (165/406)           & 69.23\% (162/234)\\
Rationale rate         & 23.0\% (328/1,426)            & 14.29\% (60/420)\\
\makecell[l]{Rationale rate of\\documented changes} & 60.63\% (328/541) & 21.43\% (60/280) \\
Rationale rate of artifacts & 50.12\% (209/417) & 15.46\% (75/485)\\
Number of rationale keypoints & 220 & 93 \\
Number of rationale codes & 70 & 33 \\
Number of rationale themes & 5 & 6 \\
\bottomrule
\multicolumn{3}{l}{\textsuperscript{$\diamond$} Computed over release lines with net growth only.} \\
\multicolumn{3}{l}{\textsuperscript{$\diamond\diamond$} Stratified random sample across change types (95\% CI, $\pm$5\% margin of error).} \\
 \\
\end{tabular}
\end{table}

\subsection{Change Frequency and Type Distribution (RQ1.1)}
\label{subsect:rq1.1}
\begin{figure}[htb]
\centering
\begin{subfigure}[t]{0.43\linewidth}
    \centering
    \includegraphics[width=\linewidth]{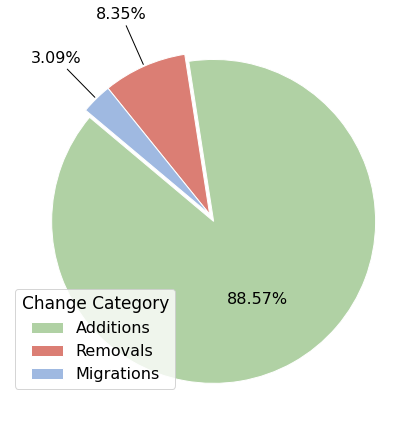}
    \caption{PTM changes}
    \label{fig:pie1}
\end{subfigure}
\hfill
\begin{subfigure}[t]{0.43\linewidth}
    \centering
    \includegraphics[width=\linewidth]{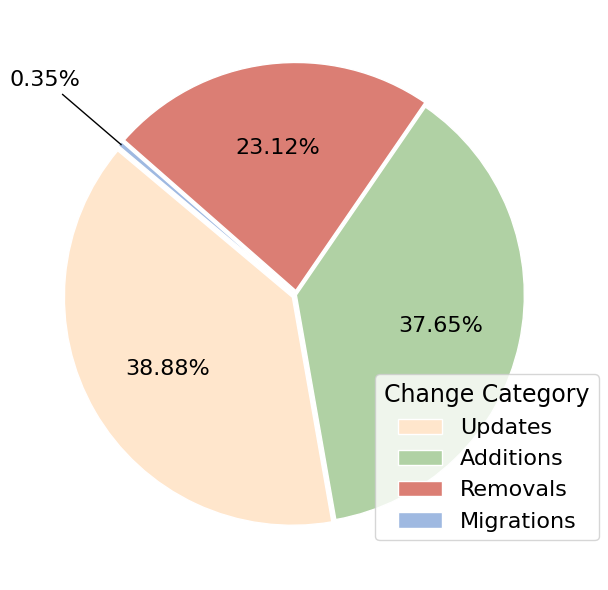}
    \caption{Library changes}
    \label{fig:pie2}
\end{subfigure}
\caption{Distribution of change types for PTMs and Libraries.}
\label{fig:pie_chart}
\end{figure}

Across the valid release pairs, only 14.4\% of all release pairs (406/2,814) exhibit any form of PTM change. Among these pairs, we observed a total of 1,426 PTM change events. As shown in Figure~\ref{fig:pie_chart}, PTM change events are overwhelmingly dominated by additions, which account for 88.57\% (1,263/1,426), while removals and migrations represent only 8.34\% (119) and 3.09\% (44), respectively. These numbers reflect the raw count of individual change occurrences. When viewed at the release-pair level, PTM additions appear in 355 release pairs, removals in 47 pairs, and migrations in 25 pairs. Compared with libraries, PTM evolution is less frequent but more addition-heavy. Library changes occur more frequently (38.1\% of release pairs) and evenly distributed across additions (37.65\%), removals (23.12\%), and updates (38.88\%), with migrations still rare (0.35\%).

\begin{figure}[!ht]
    \centering
    \includegraphics[width=\linewidth]{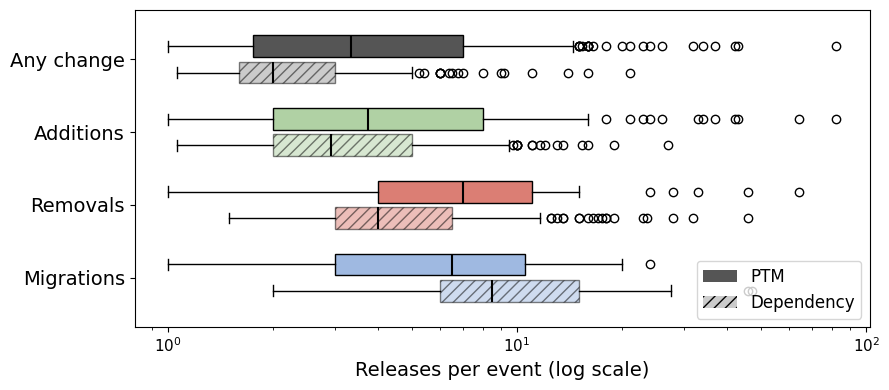}
    \caption{Comparative distribution of releases per change event for PTMs and libraies.}
    \label{fig:box_change_type}
\end{figure}

PTM changes are relatively infrequent per release pair but common over project lifetimes, affecting 65.8\% of repositories and 48.1\% of release lines. Among release lines with each PTM change type, the median line shows one addition every 3.75 releases, one removal every 7.00 releases, and one migration every 6.50 releases (Figure~\ref{fig:box_change_type}). In comparison, library changes have broader lifetime coverage and higher event cadence: 76.1\% of repositories and 56.3\% of release lines exhibit at least one library change, removals affect 35.48\% of release lines compared to only 8.20\% for PTMs, and additions/removals occur around every release (median 2.00/3.50 releases per event). While both domains show that migrations are the least frequent activity, PTMs are generally added and removed at nearly half the cadence of libraries.

To statistically confirm these cadence differences, we applied paired Wilcoxon signed-rank tests and Mann-Whitney U tests with Bonferroni correction ($\alpha$ = 0.0125), reported in Table~\ref{tab:cadence_stat}. Both tests consistently show that library changes occur significantly more frequently than PTM changes across all meaningful change types (overall, additions, and removals), all p-values $<$ 0.005 and moderate to large effect sizes (Rank-Rank-Biserial $r$ up to 0.96, Cliff's $\delta$ up to 0.45). Migration shows no significant difference in either test, likely due to limited sample size. These results confirm that the cadence gap between PTMs and libraries is consistent and robust across both within-release-line and population-level analyses.

\begin{table}[ht]
\centering
\scriptsize
\caption{Statistical comparison of PTM vs.\ library change cadence using $\alpha = 0.0125$.}
\label{tab:cadence_stat}
\renewcommand{\arraystretch}{1.0}
\begin{tabular}{l|rrrr}
\toprule
\textbf{Type} & \textbf{n} & \textbf{Medians} & \textbf{p-value} & \textbf{Effect size} \\[-0.8ex]
 & \scriptsize (PTM / Lib) & \scriptsize (PTM / Lib) & & \\
\midrule
\multicolumn{5}{l}{\textit{(a) Paired Wilcoxon Signed-Rank Test (Rank-Biserial $r$)}} \\
\midrule
Overall   & 164 / 164 & 3.88 / 1.50 & $5.28 \times 10^{-18}$\textsuperscript{***} & 0.96 \\
Addition  & 142 / 142 & 4.00 / 2.00 & $4.20 \times 10^{-11}$\textsuperscript{***} & 0.93 \\
Removal   &  28 / 28 & 8.00 / 4.00 & $3.19 \times 10^{-3}$\textsuperscript{**}  & 0.95 \\
Migration &   5 / 5 & 10.00 / 10.00 & $1.00$\textsuperscript{ns}              & 0.60 \\
\midrule
\multicolumn{5}{l}{\textit{(b) Mann-Whitney U Test  (Cliff's $\delta$)}} \\
\midrule
Overall   & 217 / 254 & 3.33 / 1.50 & $2.63 \times 10^{-17}$\textsuperscript{***} & 0.45 \\
Addition  & 200 / 223 & 3.75 / 2.00 & $1.97 \times 10^{-6}$\textsuperscript{***}  & 0.27 \\
Removal   &  37 / 160 & 7.00 / 3.50 & $7.52 \times 10^{-5}$\textsuperscript{***}  & 0.42 \\
Migration &  23 /  33 & 6.50 / 8.00 & $2.05 \times 10^{-1}$\textsuperscript{ns}   & $-$0.20 \\
\midrule
\multicolumn{5}{l}{\textsuperscript{***} $p < 0.001$; \textsuperscript{**} $p < 0.0125$; \textsuperscript{ns} not significant.} \\
\multicolumn{5}{l}{Medians = Releases per change event (higher $\Rightarrow$ less frequent).}  \\
\end{tabular}
\end{table}

\begin{summarybox}{RQ1.1}
\label{summary_rq1.1}
\noindent{\small $\bullet$} PTM changes are less frequent and tend to accumulate rather than be removed or replaced once integrated, whereas library changes are nearly three times more frequent with a more balanced type distribution. \\
\noindent{\small $\bullet$} PTMs evolve at approximately half the rate of libraries across most projects, with this difference being statistically significant for all change types except migrations.
\end{summarybox}

\subsection{Temporal Evolution of Reuse (RQ1.2)}
\label{subsect:rq1.2}
\begin{figure}[htb]
    \centering
    \includegraphics[width=\linewidth]{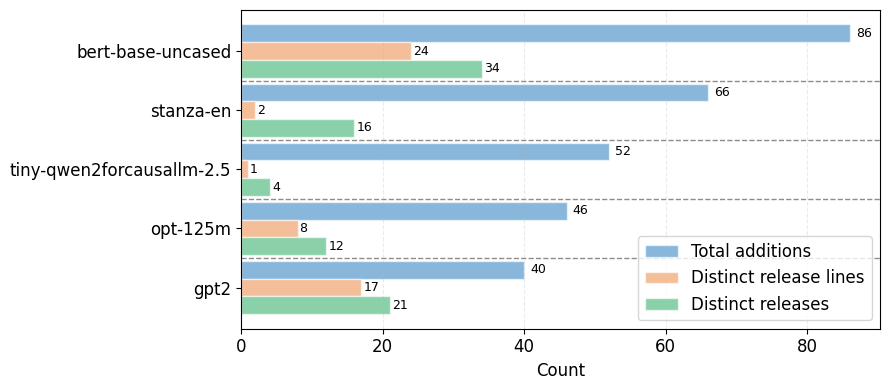}
    \caption{Top five most frequently added PTMs.}
    \label{fig:top_ptms}
\end{figure}

Given the addition-heavy PTM change pattern in RQ1.1, we next examined how PTM reuse evolves after first adoption. Across 217 release lines with PTM changes, 39.2\% show net growth, while only 10.6\% show net decline. Among growing lines, the median line ends with twice as many PTMs as at first adoption, and 41.2\% exhibit only additions with no removals or replacements, indicating purely accumulative adoption. This accumulative trend stands in stark contrast to libraries, where only 12.2\% of release lines are addition-only and the median growth factor is more incremental (1.29× vs. 2.0×). Interestingly, PTM accumulation is also reflected in the library layer. AI-focused libraries (e.g., \textit{transformers}, \textit{torch}) are frequently added alongside PTM growth, suggesting that PTM accumulation requires expanded supporting infrastructure. Figure~\ref{fig:top_ptms} shows the top five PTMs by total additions. \textit{bert-base-uncased} dominates with 86 additions across 34 release lines, while \textit{stanza} ranks second, despite appearing in only two release lines, indicating repeated accumulation within the same projects.

\begin{figure}[htb]
    \centering
    \includegraphics[width=\linewidth]{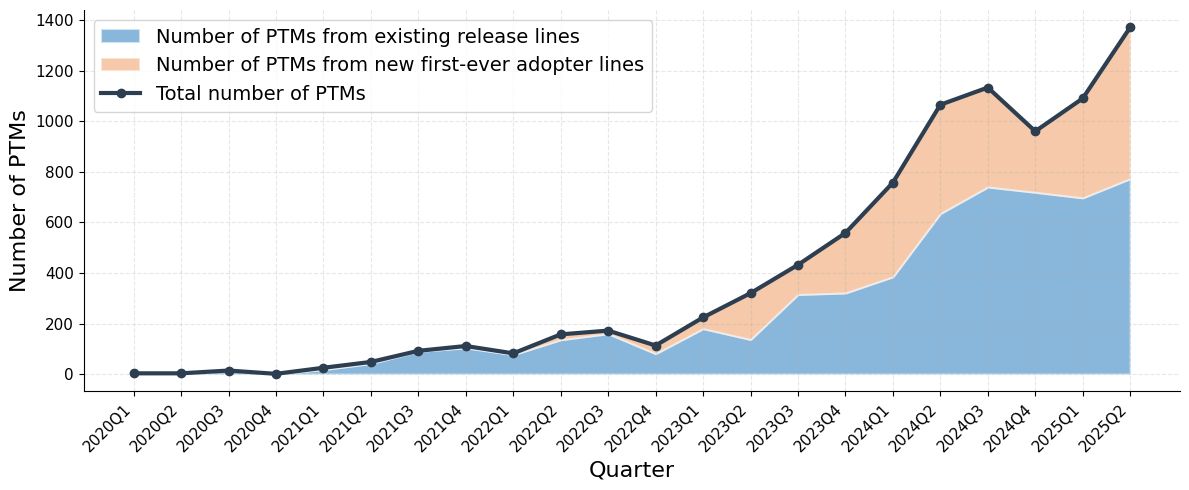}
    \caption{Quarterly number of PTMs reused, decomposed into contributions from existing adopters (blue) and first-time adopters (orange), with the total shown as a line.}
    \label{fig:timeline}
\end{figure}

To further contextualize this trend over calendar time, Figure~\ref{fig:timeline} decomposes the quarterly number of PTMs reused into contributions from projects that already use PTMs (i.e., existing adopters, shown in blue) and projects adopting PTMs for the first time (i.e., new adopters, shown in orange). While newly adopting projects contribute consistently, the dominant and growing share comes from existing projects continuing to add PTMs over time, confirming that growth in PTM reuse is primarily driven by accumulation within existing projects rather than an influx of new adopters.

\begin{figure}[htb]
    \centering
    \includegraphics[width=\columnwidth]{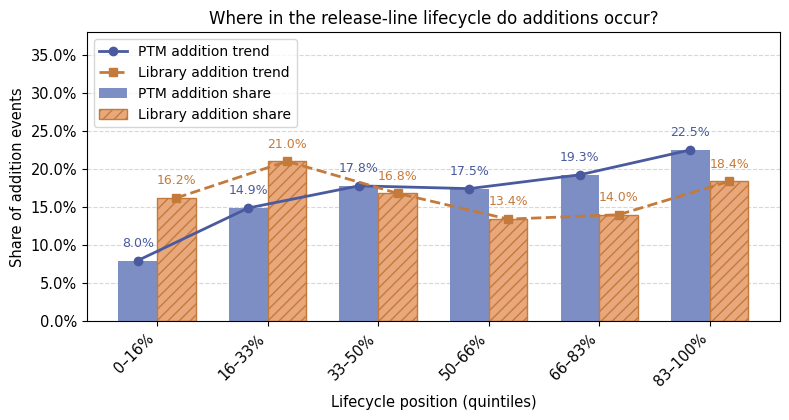}
    \caption{Comparative distribution of PTM and library addition events across the release-line lifecycle, divided into quintiles.}
    \label{fig:add_lifecycle}
\end{figure}

While the calendar-time analysis shows intensification within existing release lines, we next examined when new PTM additions occur across the release-line lifecycle. As shown in Figure~\ref{fig:add_lifecycle}, PTM additions are not uniformly distributed: early stages account for only 8.0\% of additions, while later stages rise to 19.3\% and 22.5\%, a nearly threefold increase from first to final stage. This indicates that projects tend to introduce new PTMs increasingly as they mature. In contrast, library additions peak early and dip through mid-lifecycle, further reinforcing that PTM adoption is a progressively accumulating behavior rather than a front-loaded one.

\begin{tcolorbox}[
  colback=gray!20,
  colframe=black,
  boxrule=0.8pt,
  arc=3pt,
  top=4pt, bottom=4pt, left=6pt, right=6pt,
  title={\textbf{Summary of Findings (RQ1.2)}},
  fonttitle=\bfseries,
  coltitle=black,
  colbacktitle=gray!60,
  titlerule=0.8pt,
  toptitle=2pt,
  bottomtitle=2pt
]
\label{summary_rq1.2}
\noindent{\small $\bullet$} PTM reuse grows accumulatively over project lifetimes, with most typically doubling their PTM count, a rate far exceeding the incremental growth seen in libraries. \\
\noindent{\small $\bullet$} Unlike libraries, PTM additions increase as projects mature, peaking toward the later lifecycle stages.
\end{tcolorbox}

\section{Documentation and Documented Rationales of PTM Changes (RQ2)}
\label{sect:result2}
Building on the PTM change patterns identified in RQ1, this section examines how developers document and justify PTM changes through a qualitative analysis of textual artifacts such as commit messages and pull request descriptions

\subsection{Documentation of PTM Changes (RQ2.1)}

\begin{figure}[!htb]
    \centering
    \includegraphics[width=0.9\linewidth]{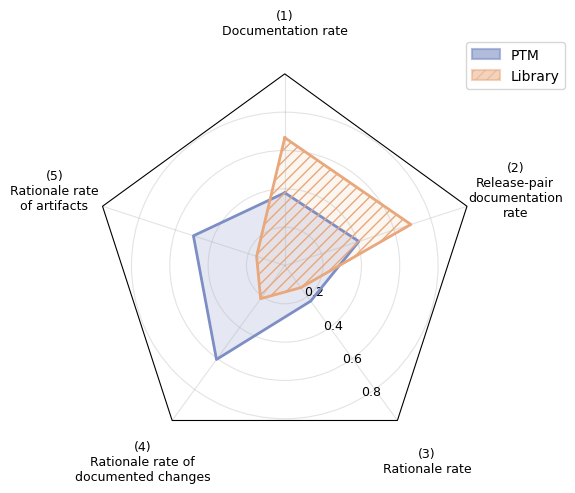}
    \caption{Documentation and rationale coverage profile comparing PTM and library changes across five metrics (Table~\ref{tab:doc_metrics}).}
    \label{fig:rq21_radar}
\end{figure}

\begin{figure}[!htb]
    \centering
    \includegraphics[width=\linewidth]{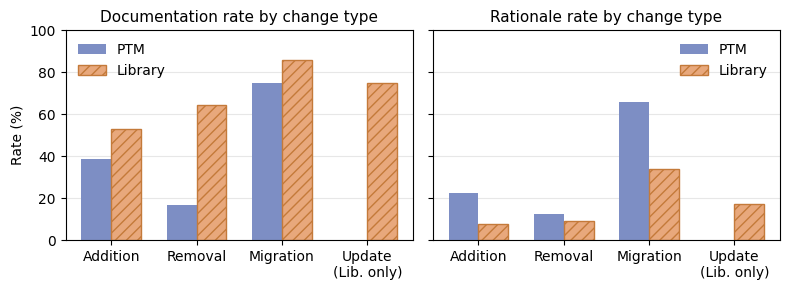}
    \caption{Documentation and rationale coverage by change type.}
    \label{fig:ptm_change_type_coverage}
\end{figure}

\begin{figure}[!htb]
    \centering
    \includegraphics[width=\columnwidth]{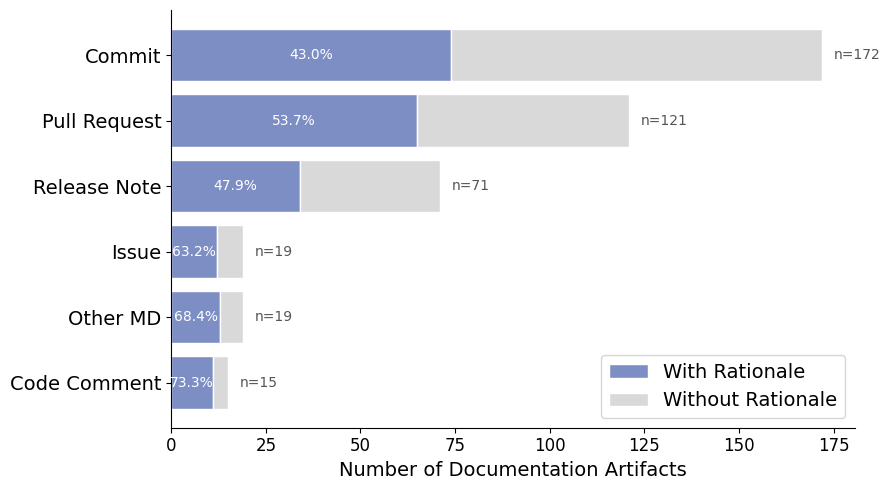}
    \caption{Distribution of documentation artifacts and rationale coverage across different artifact types for PTM changes.}
    \label{fig:doc_bar}
\end{figure}

As shown in Figure \ref{fig:rq21_radar} (metrics 1--2), fewer than half of PTM changes are explicitly documented (37.94\% of change events; 40.64\% of release pairs with changes). Documentation coverage varies substantially by change type (Figure \ref{fig:ptm_change_type_coverage}). PTM additions are documented in 38.64\% of cases and removals in only 16.81\%, while migrations are much more likely to be documented (75\%). Next, we examined where PTM change documentation appeared across different types of textual artifacts. In total, we identified 417 artifacts that documented at least one PTM change. These records live primarily in developer-centric artifacts (Figure \ref{fig:doc_bar}), led by commit messages (41.25\%) and pull request descriptions (29.02\%), with release notes appearing in only 17.03\% of cases.

Rationales are even scarcer, recorded in only 23.0\% of all PTM change events, as shown in Figure~\ref{fig:rq21_radar} (metric 3). Among the 417 identified documentation artifacts (metric 5), half contain explicit rationale information. However, when a change is documented, 60.63\% of documented changes include an explicit rationale (metric 4). Although most rationale-reported events involve PTM additions (86.59\%), only 22.49\% of all additions are documented with rationales, compared to 12.61\% for removals and 65.91\% for migrations (Figure \ref{fig:ptm_change_type_coverage}). Similar to documentation coverage, when considering only documented changes, migrations are the most consistently justified (87.88\% of documented cases), whereas additions are significantly less likely to be explained (58.20\%) despite their high volume. The distribution of these rationales across artifacts, shown in Figure \ref{fig:doc_bar}, mirrors documentation trends: commit messages and PR descriptions remain the primary sources of rationale.

\begin{tcolorbox}[
  colback=gray!20,
  colframe=black,
  boxrule=0.8pt,
  arc=3pt,
  top=4pt, bottom=4pt, left=6pt, right=6pt,
  title={\textbf{Summary of Findings (RQ2.1)}},
  fonttitle=\bfseries,
  coltitle=black,
  colbacktitle=gray!60,
  titlerule=0.8pt,
  toptitle=2pt,
  bottomtitle=2pt
]
\label{summary_rq2.1}
\noindent{\small $\bullet$} More than half of PTM changes are not explicitly documented, with removals being the least documented (16.81\%) and migrations the most (75\%). \\
\noindent{\small $\bullet$} Rationales are reported in only 23\% of PTM change events, though when documentation exists, over half include an explicit rationale, with migrations being the most consistently justified. \\
\noindent{\small $\bullet$} Developers mainly document and justify PTM changes in commit messages and pull request descriptions.
\end{tcolorbox}

\begin{table*}[t]
\scriptsize
\centering
\renewcommand*{\arraystretch}{1.2}
\caption{Taxonomy of documented rationale themes for PTM changes. For each theme and sub-theme, the table reports the number of qualitative codes (\# Codes), the raw occurrence count in documentation artifacts (Freq.), the number of PTM change events covered by the rationale with its proportion of all documented changes (\# Changes, \%), an illustrative example (Ex.), and the distribution of PTM change types (addition, removal, migration) within the theme.}
\label{tab:triggers}
\begin{tabular}{c|c|l|ccc|c|c}
\hhline{========}
\textbf{ID} & \textbf{Theme} & \multicolumn{1}{c|}{\textbf{Sub-theme}} & \textbf{\# Codes} & \textbf{Freq.} & \textbf{\# Changes (\%)} & \textbf{Ex.} & \textbf{Change Type Dist.} \\
\toprule
 &  & Need for CV functionality & 8 & 22 & 62 (18.90) & \cite{example1} & \themebar{62}{0}{0}{120} \\
 &  & Need for NLP functionality & 9 & 28 & 35 (10.67) & \cite{example2} & \themebar{35}{0}{0}{120} \\
 &  & Need for PTM development \& optimization functionality & 4 & 15 & 12 (3.66) & \cite{example3} & \themebar{12}{0}{0}{120} \\
 &  & Need for domain-specific functionality & 6 & 13 & 9 (2.74) & \cite{example4} & \themebar{9}{0}{0}{120} \\
 &  & User-requested PTM or feature support & 2 & 3 & 2 (0.61) & \cite{example5} & \themebar{2}{0}{0}{120} \\
 \cline{3-8}
\multirow{-6}{*}{\textbf{T1}} & \multirow{-6}{*}{\makecell{\textbf{New functionality}\\\textbf{expansion}}}
 & \textbf{TOTAL} & \textbf{29} & \textbf{84} & \textbf{120 (36.59)} & -- & \themebar{120}{0}{0}{120} \\
\midrule
 &  & Lack of testing for data pipeline capability & 3 & 16 & 52 (15.85) & \cite{example6} & \themebar{52}{0}{0}{107} \\
 &  & Lack of lightweight, PTM-based functional testing & 2 & 9 & 35 (10.67) & \cite{example7} & \themebar{35}{0}{0}{107} \\
 &  & Lack of PTM evaluation or benchmark & 4 & 9 & 14 (4.27) & \cite{example8} & \themebar{14}{0}{0}{107} \\
 &  & Inconsistent or poor PTM output quality & 3 & 6 & 4 (1.22) & \cite{example9} & \themebar{4}{0}{0}{107} \\
 &  & Redundant test coverage & 1 & 2 & 2 (0.61) & \cite{example10} & \themebar{0}{2}{0}{107} \\
 \cline{3-8}
\multirow{-6}{*}{\textbf{T2}} & \multirow{-6}{*}{\makecell{\textbf{Insufficient PTM evaluation}\\\textbf{\& testing safeguards}}}
 & \textbf{TOTAL} & \textbf{13} & \textbf{42} & \textbf{107 (32.62)} & -- & \themebar{105}{2}{0}{107} \\
\midrule
 &  & Test execution cost \& pressure & 3 & 12 & 25 (7.62) & \cite{example11} & \themebar{10}{1}{14}{43} \\
 &  & Need for hardware acceleration \& inference speed & 4 & 8 & 13 (3.96) & \cite{example12} & \themebar{13}{0}{0}{43} \\
 &  & Storage requirement \& runtime overhead & 2 & 4 & 6 (1.83) & \cite{example13} & \themebar{2}{0}{4}{43} \\
 \cline{3-8}
\multirow{-4}{*}{\textbf{T3}} & \multirow{-4}{*}{\makecell{\textbf{Resource \& performance}\\\textbf{pressure}}}
 & \textbf{TOTAL} & \textbf{9} & \textbf{24} & \textbf{43 (13.11)} & -- & \themebar{25}{1}{17}{43} \\
\midrule
 &  & Upstream dependency incompatibility & 4 & 6 & 13 (3.96) & \cite{example14} & \themebar{5}{0}{8}{39} \\
 &  & PTM deprecation \& functionality retirement & 2 & 10 & 12 (3.66) & \cite{example15} & \themebar{0}{12}{0}{39} \\
 &  & Upstream PTM \& dependency update & 4 & 9 & 11 (3.35) & \cite{example16} & \themebar{0}{1}{10}{39} \\
 &  & New upstream usage policy and access constraint & 2 & 5 & 3 (0.91) & \cite{example17} & \themebar{1}{0}{2}{39} \\
 \cline{3-8}
\multirow{-5}{*}{\textbf{T4}} & \multirow{-5}{*}{\makecell{\textbf{Upstream- \& lifecycle-driven}\\\textbf{changes}}}
 & \textbf{TOTAL} & \textbf{12} & \textbf{30} & \textbf{39 (11.89)} & -- & \themebar{6}{13}{20}{39} \\
\midrule
 &  & Lack of input pre-processing capability & 3 & 14 & 15 (4.57) & \cite{example18} & \themebar{15}{0}{0}{33} \\
 &  & Lack of representation learning and encoding capability & 2 & 11 & 10 (3.05) & \cite{example19} & \themebar{10}{0}{0}{33} \\
 &  & Lack of modality-specific feature extraction capability & 2 & 6 & 8 (2.44) & \cite{example20} & \themebar{8}{0}{0}{33} \\
 \cline{3-8} 
\multirow{-4}{*}{\textbf{T5}} & \multirow{-4}{*}{\makecell{\textbf{Data pipeline}\\\textbf{capability gaps}}}
 & \textbf{TOTAL} & \textbf{7} & \textbf{31} & \textbf{33 (10.06)} & -- & \themebar{33}{0}{0}{33} \\
\bottomrule
\multicolumn{8}{l}{\textit{Note:} Bars represent sub-theme contribution to the parent theme total, segmented by change type: {\color{colorAdd}$\blacksquare$} Addition {\color{colorRem}$\blacksquare$} Removal {\color{colorMig}$\blacksquare$} Migration.}
\end{tabular}
\end{table*}

\subsection{Taxonomy of PTM Change Rationales (RQ2.2)}

Table~\ref{tab:triggers} presents the taxonomy of documented rationales for PTM changes. We identified rationales from 209 documentation artifacts that explicitly explain why a PTM change occurred. The taxonomy has three levels: 70 low-level codes, grouped into 20 sub-themes, and consolidated into five high-level themes. Together, these rationales explain 328 documented PTM changes (additions: 284, removals: 15, migrations: 29). The themes are: (1) New functionality expansion, (2) Insufficient PTM evaluation \& testing safeguards, (3) Resource \& performance pressure, (4) Upstream- \& lifecycle-driven changes, and (5) Data pipeline capability.

\begin{themebox}
    \text{Theme 1: New Functionality Expansion (84)}
\end{themebox}
New functionality expansion is the most frequent theme, explaining 120 changes (36.59\%). These changes introduce new functionality or support new PTMs within existing functionality. A large portion support new application tasks in computer vision (CV) and natural language processing (NLP), including text-to-image generation, background removal, speech-to-text, and relation extraction. For instance, projects integrate PTMs such as \textit{shuttleai/shuttle-3-diffusion} for image generation (Example~\cite{example1}) or spaCy models (e.g., \textit{da\_core\_news\_sm}) for relation extraction (Example~\cite{example2}). Another group focuses on PTM additions in tooling projects, extending capabilities for PTM development, training, and optimization, including retrieval-augmented inference, mixture-of-experts architectures, parameter-efficient fine-tuning, and compression-aware training. For example, the DeepSeek model is integrated to enable MoE sparse parallelism (Example~\cite{example3}). Domain-specific adaptations also appear, covering security analysis, medical imaging, and scientific applications such as protein structure prediction. Some additions are directly driven by user requests (Example~\cite{example5}).

\begin{figure}[htb]
    \centering
    \includegraphics[width=\linewidth]{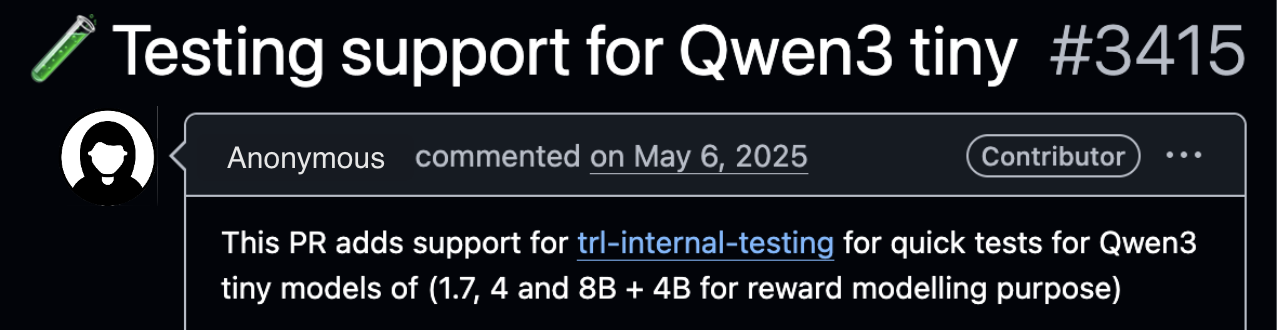}
    \caption{Example of PTM lightweight testing  (Example~\cite{example7}).}
    \label{fig:example7}
\end{figure}

\begin{themebox}
    \text{Theme 2: Insufficient PTM Evaluation \& Testing Safeguards (42)}
\end{themebox}
This theme captures documented rationales caused by gaps in PTM evaluation, testing, or validation practices. These gaps surface when projects lack adequate tests to ensure correct PTM behavior. This theme ranks second in both frequency and share of PTM changes, accounting for 107 changes (32.62\%). A major portion of these changes arises from insufficient testing of data pipelines, including preprocessing, tokenization, and multilingual compatibility. Developers often introduce additional PTMs to validate output correctness and ensure robustness across diverse inputs. For example, one project adds multilingual PTMs to test tokenization across nine African languages (Example~\cite{example6}). Another common pattern involves introducing lightweight, PTM-based tests to support regression testing while minimizing execution overhead. These tests help detect functional breakages early without incurring high computational cost. For instance, a project integrates lightweight PTM-based tests to validate PTM behavior efficiently, as shown in Figure \ref{fig:example7}.

We also observed PTM changes motivated by the lack of clear evaluation or benchmarking. In such cases, developers introduce additional PTMs to establish evaluation baselines or validate PTM behavior, such as benchmarking performance or ensuring temporal consistency (Example~\cite{example8}). In addition, some changes address inconsistent or poor output quality, where PTM outputs deviate from expected formats or semantics. These issues often lead to PTM replacement or adjustment to restore correctness (Example~\cite{example9}). A smaller number of cases involve removing redundant PTM-based tests to reduce maintenance overhead and simplify the test suite, as exemplified in Example~\cite{example10}.

\begin{themebox}
    \text{Theme 3: Resource \& Performance Pressure (24)}
\end{themebox}

\begin{figure}[htb]
\centering
\includegraphics[width=\linewidth]{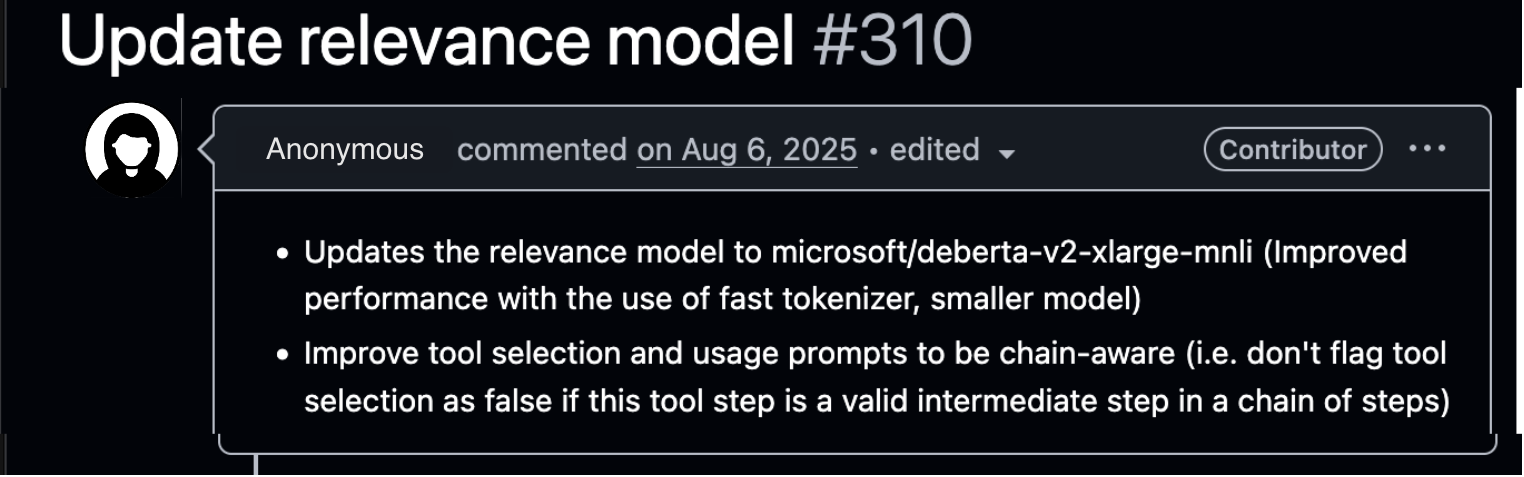}
\caption{Example of performance optimization (Example~\cite{example12}).}
\label{fig:example12}
\end{figure}

This theme captures PTM changes motivated by constraints on computational resources, execution time, or system performance, accounting for 43 changes (13.11\%). Many stem from the high cost of test execution, where developers substitute full-scale models with lightweight or randomly initialized alternatives to keep testing efficient. For example, PrunaAI replaces its original PTM with \textit{yujiepan/whisper-v3-tiny-random} to reduce execution time and alleviate CI pressure (Example~\cite{example11}). Inference performance also drives changes, where smaller or more efficient PTMs and optimized tokenizers improve latency and throughput (Figure~\ref{fig:example12}). Storage constraints further motivate replacements with lightweight alternatives, for instance, substituting \textit{whisper-small} with \textit{whisper-tiny} to reduce disk usage (Example~\cite{example13}).

\begin{figure}[htb]
\centering
\includegraphics[width=\linewidth]{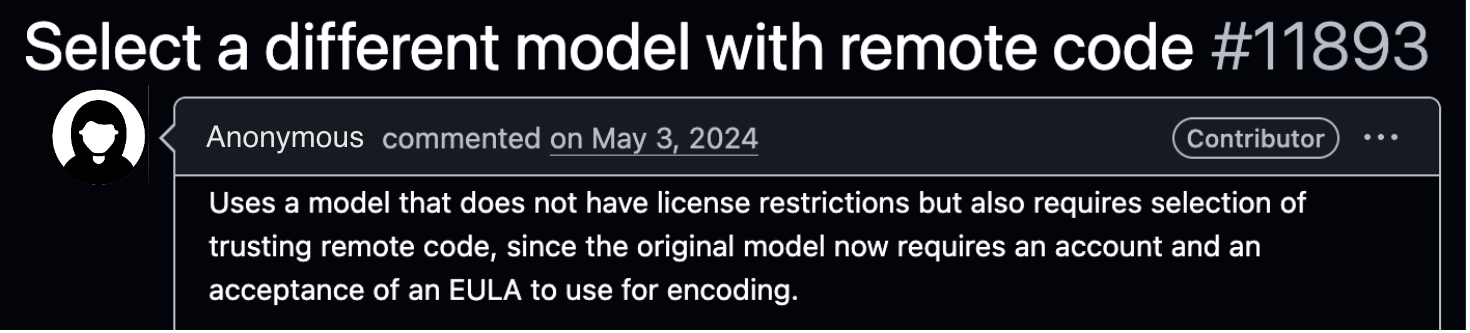}
\caption{Example of a PTM migration driven by upstream licensing restrictions (Example~\cite{example17}).}
\label{fig:example17}
\end{figure}

\begin{themebox}
    \text{Theme 4: Upstream- and Lifecycle-Driven Changes (30)}
\end{themebox}
This theme covers external forces from upstream PTMs, libraries, or policies that require downstream adaptation even without direct user demand, explaining 39 PTM changes (11.89\%). A major driver is upstream technical evolution---dependency incompatibilities, model updates, or library changes that break existing integrations. For example, a project removes an older \textit{convnext} model after encountering performance issues with a newer \textit{transformers} version (Example~\cite{example16}). Changes in upstream usage policies, licensing terms, or access restrictions also prompt migrations, for instance, one developer switched to \textit{gte-base-en-v1.5} to avoid EULA requirements and account-gated restrictions of the previous PTM (Figure~\ref{fig:example17}). Finally, some changes reflect downstream decisions to retire PTMs or discontinue associated functionality due to lack of user demand (Example~\cite{example15}).

\begin{themebox}
    \text{Theme 5: Data Pipeline Capability Gaps (31)}
\end{themebox}

This documented rationales explain 33 changes (10.06\%), the smallest share among themes. One group addresses insufficient input handling, where PTMs are added for tokenization and prompt construction to process diverse inputs—for example, integrating \textit{bert-base-uncased} as a WordPiece tokenizer for document and query processing (Example~\cite{example18}). Another group targets representation capability, integrating PTMs for modality-specific feature extraction and semantic encoding to generate richer embeddings for retrieval, matching, and reasoning tasks (Examples~\cite{example19, example20}).

\begin{tcolorbox}[
  colback=gray!20,
  colframe=black,
  boxrule=0.8pt,
  arc=3pt,
  top=4pt, bottom=4pt, left=6pt, right=6pt,
  title={\textbf{Summary of Findings (RQ2.2)}},
  fonttitle=\bfseries,
  coltitle=black,
  colbacktitle=gray!60,
  titlerule=0.8pt,
  toptitle=2pt,
  bottomtitle=2pt
]
\label{summary_rq2.2}
\noindent{\small $\bullet$} New functionality expansion is the most frequently discussed rationale, and also covers the largest share of documented PTM changes (36.59\%). \\
\noindent{\small $\bullet$} Insufficient testing safeguards are the second most impactful rationale (32.62\%), highlighting that PTM correctness validation is significant for PTM reuse. \\
\noindent{\small $\bullet$} Resource and performance pressure drives most PTM migrations (58.6\%), indicating that replacement is the primary response to computational constraints. \\
\noindent{\small $\bullet$} Upstream and lifecycle drivers show strong change-type alignment: PTM deprecation exclusively drives removals, while upstream updates predominantly trigger migrations. \\
\noindent{\small $\bullet$} Data pipeline gaps exclusively drive PTM additions, suggesting that such gaps are always addressed by expanding PTM support rather than replacing existing ones.
\end{tcolorbox}

\begin{table*}[t]
\scriptsize
\centering
\renewcommand*{\arraystretch}{1.2}
\caption{High-level themes of documented rationales for library changes. For each library theme, the table reports the overlapping PTM theme(s), overlapping prior work (\cite{Teyton2014, Kabinna2016, He2021, Barbosa2022, Jaisri2025, Bavota2015}), the raw occurrence count in documentation artifacts (Freq.), the number of library change events covered by the theme with its proportion of all library changes with rationales (\# Lib. Changes, \%), and the distribution of library change types (addition, removal, migration, version update).}
\label{tab:dep_trigger_themes}
\begin{tabular}{l|c|c|cc|c}
\hhline{======}
\textbf{Theme} & \textbf{Overlapping PTM Theme ID} & \textbf{Overlapping Prior Work} & \textbf{Freq.} & \textbf{\# Lib. Changes (\%)} & \textbf{Change Type Dist.} \\
\toprule
Compatibility \& integration issues & T4 & \cite{He2021, Teyton2014, Kabinna2016, Barbosa2022, Bavota2015} & 42 & 37 (61.67) & \depthemebar{7}{1}{7}{22}{37} \\
Upstream quality, maturity, \& evolution & T4 & \cite{He2021, Teyton2014, Jaisri2025, Bavota2015} & 19 & 17 (28.33) & \depthemebar{1}{4}{12}{0}{17} \\
Functionality \& capability needs & \makecell{T1, T5} & \cite{He2021, Teyton2014, Kabinna2016, Bavota2015} & 14 & 13 (21.67) & \depthemebar{7}{0}{4}{2}{13} \\
Maintainability \& simplification & -- & \cite{He2021, Teyton2014, Barbosa2022, Jaisri2025} & 9 & 6 (10.00) & \depthemebar{1}{4}{1}{0}{6} \\
Performance \& footprint pressure & T3 & \cite{He2021, Teyton2014, Kabinna2016, Jaisri2025} & 5 & 5 (8.33) & \depthemebar{1}{2}{1}{1}{5} \\
Security \& policy constraints & T4 & \cite{He2021, Jaisri2025, Bavota2015} & 4 & 3 (5.00) & \depthemebar{0}{0}{3}{0}{3} \\
\cline{1-6}
\textbf{TOTAL} & -- & -- & \textbf{93} & \textbf{60 (100.00)} & \depthemebar{11}{8}{17}{24}{60} \\
\bottomrule
\multicolumn{6}{l}{\textit{Notes:} Bars show change type distribution: {\color{colorAdd}$\blacksquare$} Addition {\color{colorRem}$\blacksquare$} Removal {\color{colorMig}$\blacksquare$} Migration {\color{colorUpd}$\blacksquare$} Update.}
\end{tabular}
\end{table*}

\subsection{PTM vs Library Documentation and Rationales (RQ2.3)}
\label{subsect:rq2.3}
Revisiting Figure~\ref{fig:rq21_radar} from a comparative perspective, PTM and library changes show different coverage profiles. Library changes are documented more often at both the event and release levels, while PTM changes are more likely to include explicit rationales once documented (60.63\% vs. 21.43\%). At the artifact level, rationale reporting is also much denser for PTM changes. This suggests library changes are communicated more routinely, whereas PTM changes, when documented, are more frequently accompanied by rationale. A similar contrast appears at the change-type level (Figure~\ref{fig:ptm_change_type_coverage}). For every change type, library changes show higher documentation coverage than PTM changes, with the largest gap in removals. In contrast, rationale coverage shows the same pattern in both domains, highlighting that migrations are most often justified.

To support cross-domain comparison of documented rationales, Table~\ref{tab:dep_trigger_themes} reports the resulting high-level library rationale themes and their links to overlapping PTM themes. Library changes are dominated by \textit{Compatibility \& integration issues} and \textit{Upstream quality, maturity, \& evolution}. These drivers primarily lead to migrations and updates, reflecting a reactive effort to keep the system functional as the external ecosystem shifts. In contrast, the leading PTM rationales are new functionality and capability expansion (\textbf{T1, T2}), which together account for nearly half of all PTM changes. Unlike libraries, which are often updated to fix broken integrations, PTMs are added to unlock new capabilities. Even where functionality drivers (\textbf{T1}, \textbf{T5}) overlap with library needs, the connection is only partial in high level; PTM changes remain fundamentally model-centric, rather than general software utility.

A striking difference is the role of PTM testing (\textbf{T2}). This theme accounts for nearly a third of PTM changes but has no direct equivalent in our library taxonomy or prior literature. In traditional software, developers rarely add a second library just to verify the first. However, developers frequently introduce alternative PTMs to handle unstable outputs, provide benchmarks, or act as lightweight testing surrogates. Even when overlap exists (e.g., \textbf{T3} with performance/footprint, \textbf{T4} with upstream constraints), documented PTM rationales remain tied to PTM behavior and inference context, not only package management constraints. Moreover, library rationales, even within \textit{Functionality \& capability needs}, are spread across multiple change types, particularly migrations and updates. Conversely, documented rationales for PTM functionality (\textbf{T1}, \textbf{T5}) and evaluation (\textbf{T2}) lead almost exclusively to additions.

\begin{tcolorbox}[
  colback=gray!20,
  colframe=black,
  boxrule=0.8pt,
  arc=3pt,
  top=4pt, bottom=4pt, left=6pt, right=6pt,
  title={\textbf{Summary of Findings (RQ2.3):}},
  fonttitle=\bfseries,
  coltitle=black,
  colbacktitle=gray!60,
  titlerule=0.8pt,
  toptitle=2pt,
  bottomtitle=2pt
]
\label{summary_rq2.3}
\noindent{\small $\bullet$} Library changes are more consistently documented than PTM changes, but when PTM changes are documented, they are more likely to include an explicit rationale. \\
\noindent{\small $\bullet$} PTM changes are primarily driven by capability expansion, whereas library changes are mainly motivated by ecosystem maintenance and compatibility. \\
\noindent{\small $\bullet$} PTM changes exhibit a unique rationale: insufficient evaluation and testing safeguards, where developers introduce additional PTMs to validate or surrogate-test existing ones, reflecting the uncertainty of PTM behavior.
\end{tcolorbox}

\section{Discussions \& Implications}
\label{sect:discussions}
\subsection{Challenges in Identifying PTM Migration and Update}
\label{subsect:challenge_migration}

Unlike traditional libraries, PTMs rarely have clear functional equivalents. While developers can often recognize interchangeable libraries (e.g., \textit{requests} vs.\ \textit{httpx} \cite{Islam2024}), PTMs do not share this property. Even PTMs within the same family may serve different downstream tasks or behave differently in practice. More importantly, the same PTM identifier can fulfill multiple roles within a system. As shown in Listing~\ref{lst:ptm-multi-role}, a PTM can serve simultaneously as a tokenizer and a task-specific predictor \cite{meleksabit2025example}. This role-dependent reuse means that a change involving the same PTM may reflect a shift in preprocessing or inference logic rather than a straightforward replacement, making it difficult to infer ``which PTM changes to which'' using name matching or semantic similarity alone.

\begin{lstlisting}[style=ptmcode, language=Python, float=htb, floatplacement=htb, caption={The same PTM identifier can serve multiple roles in a downstream pipeline (source:~\cite{meleksabit2025example}).}, label={lst:ptm-multi-role}]
tokenizer = AutoTokenizer.from_pretrained(
    "distilbert-base-uncased-finetuned-sst-2-english")
model = AutoModelForSequenceClassification.from_pretrained(
    "distilbert-base-uncased-finetuned-sst-2-english")
classifier = pipeline("text-classification", model=model, tokenizer=tokenizer)
\end{lstlisting}

Prior work on library migration assumes that replacements can be predefined based on functional equivalence \cite{Teyton2014, Kabinna2016, Barbosa2022} and tracked through dependency file changes \cite{He2021, Islam2023}. This assumption does not hold for PTMs. Beyond role-dependent reuse, PTM metadata provides limited support for reliable pairing: model-card fields describing functionality or lineage (e.g., variant types) are frequently missing, inconsistently specified across configuration files and tags, or too coarse, and many PTMs carry multiple user-defined or ambiguous tags \cite{Ajibode2025}. The challenge is further compounded by the absence of standardized PTM versioning~\cite{Ajibode2025}. In our library baseline, version updates account for 38.88\% of all changes, yet this change type is effectively invisible in the PTM domain. Although recent work suggests inferring PTM versions from naming conventions, more than 93\% of Hugging Face PTMs do not explicitly encode version information in their names~\cite{Ajibode2025}. We observed the same in our dataset: across 539 unique PTMs spanning 7,604 file snapshots, only 15.9\% (86) contained any version indicator. Moreover, only four PTM families contain more than one version, covering just eight PTMs in total. Of these, only four appear within the same repositories, limiting their usefulness for identifying within-project PTM updates.

Finally, although PTMs hosted on Hugging Face are stored in versioned Git repositories and developers can pin a specific PTM version via the \texttt{revision} parameter in \texttt{from\_pretrained()}, explicit use of this mechanism remains rare: only 12\% of projects specify a PTM revision~\cite{Yasmin2025}, a limitation we observe equally in our dataset. Together, these challenges indicate that migration and update detection approaches developed for libraries are inapplicable to PTMs.

\subsection{Changes in PTM Reuse vs Library Reuse}

\subsubsection{\textbf{Frequency and Pattern of Change}}
Both PTMs and libraries change infrequently, consistent with prior work~\cite{Zaimi2015, Kula2018}. However, their change profiles diverge in a fundamental way: while traditional libraries show balanced addition, removal, and version update activity, PTM changes are overwhelmingly addition-dominated, with removals and migrations remaining comparatively rare, as summarized in RQ1.1. Prior studies report addition rates of only 3\%--9\% but removal rates of 30\%--41\% across ecosystems~\cite{Zaimi2015, He2021, Jaisri2025}, a pattern essentially inverted in the PTM domain. This cadence gap is statistically robust, with PTMs being added or removed at roughly half the pace of traditional libraries (see RQ1.1).

We interpret this divergence as a shift from \emph{substitution to accumulation.} Library reuse often follows what we term ``vertical growth'' (i.e., upgrade or replacement within a defined role). In AI-enabled systems, PTMs often coexist. Because PTMs are flexible and multi-purpose, a developer might add a new PTM for a specific task (such as multilingual support) or to accommodate differing performance or resource trade-offs, without removing existing ones. This leads to the ``horizontal expansion'' we observe, where the PTM count grows progressively as the project matures (See~RQ1.2), unlike libraries, whose changes show no systematic lifecycle trend~\cite{Terzi2022}.

Notably, this PTM accumulation seems to drive a secondary growth in infrastructure. We found that as PTM reuse grows, specialized libraries like \textit{transformers} are among the most frequently added, suggesting that adopting more PTMs necessitates a corresponding growth in the supporting dependency layer. This may partly explain why library additions account for the largest share of library changes in our baseline (37.65\%), a notably higher addition rate than reported in prior library studies. Collectively, this progressive accumulation could increase the number of coexisting PTMs and their associated dependencies, tests, and configurations, possibly expanding the system's maintenance surface over time.

\subsubsection{\textbf{Documentation and Rationales}}

The reasons for changing PTMs reveal a fundamental shift in developer priorities. While library changes are usually ``reactive'', driven by the need to fix bugs, update security, or maintain compatibility~\cite{He2021}, PTM changes are overwhelmingly ``proactive'' (see RQ2.3). We interpret this as a transition from \emph{a maintenance mindset to a capability mindset.} In traditional development, a developer changes a library because they \emph{must} (e.g., the old version is deprecated, contains bugs, or has become incompatible with the environment). Prior studies confirm that these library changes are largely reactive, driven by performance, compatibility, and security concerns~\cite{He2021, Jaisri2025, Kabinna2016, Pashchenko2020, Huang2022, Bavota2015}. In contrast, a developer adds a PTM because they \emph{aim} to (e.g., to add a new capability or improve accuracy). This explains why new functionalities are the primary drivers for PTMs (Themes~1 and~5 in RQ2.2), directly reinforcing the accumulative, horizontal growth pattern identified in RQ1 and the expansion needs reported in our prior work~\cite{banyongrakkul2025}.

This difference in documentation style suggests that developers handle \emph{PTM uncertainty} (i.e., the difficulty of predicting whether a PTM will behave correctly for a given task or dataset) differently than library stability. While library documentation is often routine, PTM documentation, when present, tends to be more rationale-heavy and intended for peer developers (as summarized in RQ2.3). This likely reflects the non-deterministic nature of PTMs; unlike traditional libraries, whose outputs are largely deterministic, a PTM’s correctness is often subjective and data-dependent, requiring developers to justify the selection and behavior of a PTM rather than just its integration. To improve transparency, developers should invest more effort in documenting PTM-related changes systematically, aligning with the practices suggested by \cite{Gao2026b}.

This unique uncertainty appears to have fostered a behavior we term ``Testing by Proxy'' (i.e., adding a new PTM specifically to benchmark or validate an existing one rather than for direct production use), as observed in RQ2.3, largely driven by Theme 2 in RQ2.2. Our data suggests that developers often add a new PTM not only for production use, but also to benchmark or validate existing ones. This reflects a fundamental uncertainty in PTM behavior that has no parallel in conventional library reuse. However, this accumulative behavior sometimes eventually hits practical limits. As shown in Theme 3 in RQ2.2, computation pressure predominantly drive PTM migrations, forcing a transition from expansion to optimization. Furthermore, functionality deprecation exclusively drives removals, while upstream updates predominantly drive migrations (Theme 4 in RQ2.2).

\subsection{Implications}
We translate our findings into practical implications for researchers and practitioners, bridging the gap between traditional dependency management and PTM evolution.

\subsubsection{\textbf{Researchers}}
{\large \textcircled{\normalsize 1}} \textbf{Managing the PTM Portfolio.} PTM reuse behaves more like portfolio expansion than substitution (RQ1.2), researchers could explore portfolio management frameworks that account for PTM coexistence, resource budgeting, and long-term maintenance, rather than focusing solely on version update tools. This includes creating impact-aware visualizations that map how a single PTM connects to specific data pipelines, test cases, and code references over time. {\large \textcircled{\normalsize 2}} \textbf{Identifying PTM Bloat and Shadow Debt.} PTM additions accumulate with rare removals (RQ1.1, RQ1.2), partly motivated by continuous capability expansion (Themes 1, 5 in RQ2.2) and task diversification (Section~\ref{subsect:challenge_migration}), raising a risk of PTM bloat~\cite{Soto-Valero2021}. Orphaned PTM references may stay hidden in configurations or test suites long after a PTM is no longer used. Empirical studies and static analysis tools could help to detect and manage such stale references before they accumulate into hidden technical debt. {\large \textcircled{\normalsize 3}} \textbf{Defining PTM Evolution Smells.} Analogous to code smells \cite{Giordano2023, Costal2024}, certain PTM evolution patterns may signal poor engineering practices, for example, rapid unchecked accumulation, PTMs with no associated tests, or additions with no documented rationale (RQ2.1). Future work could explore defining and cataloging such PTM evolution smells, and investigate the feasibility of automatic detection tools to help developers identify risky reuse patterns early. {\large \textcircled{\normalsize 4}} \textbf{Formalizing ``Testing by Proxy.''} The fact that developers add PTMs specifically to validate others (Theme 2 in RQ2.2, RQ2.3) highlights a gap in PTM behavior verification. Future work could empirically characterize how developers use surrogate PTMs for validation, then build on these findings to develop lightweight testing frameworks and standardized validation benchmarks~\cite{Zhang2022}.

\subsubsection{\textbf{Downstream Developers}}
{\large \textcircled{\normalsize 1}} \textbf{Planning for Infrastructure Scaling.} PTM reuse tends to grow twofold over project lifetimes, as reported in RQ1.2, driven by capability expansion (Themes 1, 5 in RQ2.2). Developers may benefit from anticipating the corresponding expansion in AI-specialized infrastructure and proactively managing storage, runtime, and CI overhead (Theme 3 in RQ2.2), consistent with proactive infrastructure monitoring practices in traditional software engineering~\cite{Jantti2017}. {\large \textcircled{\normalsize 2}} \textbf{Implementing PTM Inventory Management.} Developers may also adopt an explicit registry, analogous to a Software Bill of Materials (SBOM) \cite{Nocera2025}, to track which PTMs are active, their specific roles (e.g., tokenizer vs. classifier), and their validation history. This inventory would be critical for impact analysis and identifying dead PTMs before they become hidden technical debt. {\large \textcircled{\normalsize 3}} \textbf{Standardizing PTM Change Rationale and Deprecation Signaling.} Removals and replacements are significantly less documented than additions (RQ2.1), yet they may carry the highest maintenance risk. Developers should record the rationale for PTM removals in commit messages or pull requests, where most PTM documentation already concentrates, to support long-term reproducibility and maintenance~\cite{Gao2026}. Especially when retiring PTM-based functionality entirely (Theme 4 in RQ2.2), they should also explicitly communicate this to users (e.g., via release notes) so that downstream consumers can adapt in time.

\subsubsection{\textbf{Upstream PTM Hub Providers}}
{\large \textcircled{\normalsize 1}} \textbf{Enforcing Metadata and Versioning Standards.} Due to the PTM migration and versioning challenges (Section~\ref{subsect:challenge_migration}), hub providers could consider requiring standardized schemas and versioning conventions at upload time~\cite{Leipzig2021} to support downstream tracking and lineage, echoing calls for standardized ML model metadata~\cite{Gesese2025}. {\large \textcircled{\normalsize 2}} \textbf{Providing CI-Friendly Evaluation Assets.} Since testing gaps and CI costs frequently drive PTM changes (Themes 2, 3 in RQ2.2), hubs may offer lite versions of hosted PTMs. Providing standardized, randomly initialized tiny checkpoints and deterministic sample I/O pairs would allow developers to validate their pipelines without the high computational overhead that currently triggers PTM churn. {\large \textcircled{\normalsize 3}} \textbf{Automating Lifecycle and Policy Transparency.} Since upstream model updates and access policy changes drive downstream PTM migrations (Theme 4 in RQ2.2), hub providers could consider exposing structured, machine-readable metadata covering licensing terms (e.g., SBOM/SPDX)~\cite{Nocera2025b}, known dependency constraints, and deprecation timelines. Such metadata could help downstream tools surface compatibility risks and policy changes earlier, potentially reducing reactive churn.


\section{Threats to Validity}
\label{sect:threats}
\textbf{External Validity:} Our findings are based on OSS repositories reusing open-source PTMs from Hugging Face and may not generalize to closed-source or industrial systems, or to domains using different PTM types. Our PTM changes and library baseline are constructed from Python-based projects. Findings may differ in other language ecosystems. Furthermore, our release-based filtering criteria favor more mature and actively maintained projects; therefore, our findings may better reflect well-maintained, actively released PTM-based projects rather than the broader ecosystem, and caution should be exercised when generalizing to smaller or less active projects. Additionally, our stratified sample of 420 library change events, while statistically representative at 95\% confidence with $\pm$5\% margin of error, may not fully capture the diversity of motivations across all project types.

\textbf{Internal Validity:} Our migration detection relies on file-level co-occurrence of additions and removals, which may miss migrations spanning multiple commits or files. Our cross-domain comparison also assumes release pairs are comparable across domains, though confounding factors such as project size, age, or domain may influence observed differences. To partially mitigate this, all analyses were applied to the identical set of release pairs, ensuring differences reflect artifact-level behavior rather than sampling bias. A further threat relates to our statistical comparison of change cadence. Non-parametric tests assume independent observations, but release lines from the same repository may be correlated. To mitigate this, we use paired analysis where applicable, complement it with population-level tests, and apply Bonferroni correction across change types. As these types are not fully independent, the correction may be conservative; however, our results remain significant even under this stricter threshold.

Another threat arises from temporal asymmetry between PTM and library lifecycles, as traditional libraries often predate first PTM adoption.  Analyzing the full repository timeline would therefore bias library counts due to a longer observation window. To mitigate this, we anchor both PTM and library change detection to $t_1$, the point of first PTM adoption, treating this as the inception of the project as an AI-enabled system. While this ensures a fair and comparable observation, it may undercount library changes before $t_1$ and limit generalizability. Additionally, our six-stage lifecycle normalization and minimum six-release threshold are design choices that balance granularity with data availability, but alternative parameterizations could yield different distributional patterns.

\textbf{Construct Validity:} Our repository–PTM mapping and snapshot extraction rely on static code analysis, capturing PTM names specified as hard-coded strings or user-defined variables, but may miss PTMs defined indirectly via configuration files (e.g., JSON files), command-line arguments, function parameters, runtime inputs, local file paths, or identifiers absent at parse time. Additionally, our cadence metric (i.e., releases per change event) treats all releases as equivalent units of development activity, which may not hold across projects with varying release frequencies and significance, though median-based statistics and non-parametric tests partially mitigate this.

Our qualitative analysis of documentation involves manual judgment and is subject to subjectivity~\cite{Ratner2002}. To mitigate this, two authors were involved and interrater agreement was measured. We also acknowledge that identifying rationales for changes can be challenging, particularly in long or context-rich artifacts (e.g., release notes). Exhaustively reviewing entire artifacts may introduce inconsistency and reduce replicability. To mitigate this, we adopted a structured, time-bounded procedure using keyword-based filtering to limit manual review scope. Furthermore, our analysis captures only explicitly documented motivations; implicit or undocumented reasons for changes remain outside the scope and may lead to underestimation of certain rationales. Finally, our library rationale taxonomy is derived from a stratified sample of 420 change events, as annotating the full population of detected changes is impractical. While we increased the sample beyond the statistical minimum to promote coding saturation, the sampled rationales may not fully reflect all motivations in the dataset. To mitigate this, we map our taxonomy to existing rationale themes from prior work~\cite{Teyton2014, Kabinna2016, He2021, Barbosa2022, Jaisri2025, Bavota2015}, grounding our taxonomy in a broader body of evidence.

\section{Conclusion}
\label{sect:conclusion}
Pre-trained models (PTMs) are increasingly reused in downstream OSS systems, yet their long-term evolution and maintenance characteristics remain underexplored from a software engineering perspective. In this paper, we conducted an empirical study of 3,265 releases across 451 release lines from 310 repositories to understand how PTMs are adopted, evolved, documented, and what motivates their changes, comparing them against traditional software libraries as a baseline.

Our findings reveal that unlike libraries, which evolve through reactive versioning and replacement, PTMs follow an accumulative trajectory, treated as capability-extending assets that grow horizontally over a project's lifetime. The prevalence of additions, coupled with the unique emergence of PTM testing, suggests that reuse is governed by behavioral uncertainty, further reflected in a documentation gap where PTM changes, though less frequently recorded, carry richer rationale when documented.

Ultimately, these results characterize PTMs as cumulative knowledge assets, fundamentally different from traditional software libraries. This distinction could challenge the adequacy of current dependency management tools and documentation norms. As AI-enabled systems continue to scale, researchers and practitioners may benefit from shifting away from a versioning-focused maintenance mindset toward a portfolio mindset that accounts for PTM coexistence, resource budgeting, and the long-term risks of accumulation.

\bibliographystyle{IEEEtran}
\bibliography{references}

@article{Jantti2017,
author = {J{\"{a}}ntti, Marko and Cater-Steel, Aileen},
doi = {10.4301/s1807-17752017000200004},
journal = {Journal of Information Systems and Technology Management},
number = {2},
pages = {191--218},
title = {{Proactive Management of IT Operations to Improve IT Services}},
volume = {14},
year = {2017}
}

@inproceedings{Gao24,
author = {Gao, Haoyu and Zahedi, Mansooreh and Treude, Christoph and Rosenstock, Sarita and Cheong, Marc},
title = {Documenting Ethical Considerations in Open Source AI Models},
year = {2024},
isbn = {9798400710476},
publisher = {ACM},
address = {New York, NY, USA},
doi = {10.1145/3674805.3686679},
booktitle = {Proceedings of the 18th ACM/IEEE International Symposium on Empirical Software Engineering and Measurement (ESEM 2024)},
pages = {177–188},
numpages = {12},
location = {Barcelona, Spain},
}

@article{DeepSeek-AI2025,
author = {Guo, Daya and Yang, Dejian and Zhang, Haowei and others},
doi = {10.1038/s41586-025-09422-z},
issn = {14764687},
journal = {Springer Nature},
number = {8081},
pages = {633--638},
pmid = {40962978},
address   = {London, UK},
title = {{DeepSeek-R1 incentivizes reasoning in LLMs through reinforcement learning}},
volume = {645},
year = {2025}
}

@inproceedings{Jiang2023,
archivePrefix = {arXiv},
arxivId = {2303.02552},
author = {Jiang, Wenxin and others},
booktitle = {Proceedings of the 45th International Conference on Software Engineering (ICSE 2023)},
doi = {10.1109/ICSE48619.2023.00206},
eprint = {2303.02552},
isbn = {9781665457019},
issn = {02705257},
pages = {2463--2475},
publisher = {IEEE},
address = {Piscataway, NJ, USA},
title = {{An Empirical Study of Pre-Trained Model Reuse in the Hugging Face Deep Learning Model Registry}},
year = {2023}
}

@article{Cox2019,
author = {Cox, Russ},
doi = {10.1145/3329781.3344149},
issn = {1542-7730},
journal = {Queue},
number = {2},
pages = {24--47},
title = {{Surviving Software Dependencies}},
volume = {17},
publisher = {ACM},
address   = {New York, NY, USA},
year = {2019}
}

@article{Decan2019,
author = {Decan, Alexandre and Mens, Tom and Grosjean, Philippe},
doi = {10.1007/s10664-017-9589-y},
issn = {1573-7616},
journal = {Empirical Software Engineering},
number = {1},
pages = {381--416},
title = {{An empirical comparison of dependency network evolution in seven software packaging ecosystems}},
volume = {24},
year = {2019},
publisher = {Springer},
address   = {Cham, Switzerland},
}

@article{Frakes2005,
  author = {Frakes, W. B. and Kang, Kyo},
  journal={IEEE Trans. Softw. Eng.}, 
  title={{Software reuse research: status and future}}, 
  year={2005},
  volume={31},
  number={7},
  publisher = {IEEE},
  address   = {Piscataway, NJ, USA},
  pages={529-536},
  doi={10.1109/TSE.2005.85}
}

@article{Kula2018,
archivePrefix = {arXiv},
arxivId = {1709.04621},
author = {Kula, Raula Gaikovina and German, Daniel M. and Ouni, Ali and Ishio, Takashi and Inoue, Katsuro},
doi = {10.1007/s10664-017-9521-5},
eprint = {1709.04621},
issn = {15737616},
journal = {Empirical Software Engineering},
number = {1},
pages = {384--417},
title = {{Do developers update their library dependencies?: An empirical study on the impact of security advisories on library migration}},
volume = {23},
publisher = {Springer Nature},
address   = {Cham, Switzerland},
year = {2018}
}

@article{Zhou2023,
arxivId = {2302.09419},
author = {Zhou, Ce and others},
doi = {10.1007/s13042-024-02443-6},
eprint = {2302.09419},
isbn = {1304202402},
issn = {1868808X},
journal = {International Journal of Machine Learning and Cybernetics},
publisher = {Springer},
address   = {Berlin/Heidelberg, Germany},
title = {{A Comprehensive Survey on Pretrained Foundation Models: A History from BERT to ChatGPT}},
year = {2023}
}

@inproceedings{Pashchenko2020,
author = {Pashchenko, Ivan and Vu, Duc Ly and Massacci, Fabio},
doi = {10.1145/3372297.3417232},
isbn = {9781450370899},
issn = {15437221},
booktitle = {Proceedings of the 2020 ACM SIGSAC Conference on Computer and Communications Security (CCS 2020)},
pages = {1513--1531},
title = {{A Qualitative Study of Dependency Management and Its Security Implications}},
publisher = {ACM},
year = {2020}
}

@inproceedings{Zerouali2018,
author = {Zerouali, Ahmed and Constantinou, Eleni and Mens, Tom and others},
doi = {10.1007/978-3-319-90421-4_6},
isbn = {9783319904207},
issn = {16113349},
booktitle = {Proceedings of the 17th International Conference on Software Reuse (ICSR 2018)},
series    = {Lecture Notes in Computer Science},
volume    = {10826},
pages     = {95--110},
publisher = {Springer},
address   = {Cham, Switzerland},
title = {{An Empirical Analysis of Technical Lag in npm Package Dependencies}},
year = {2018}
}

@inproceedings{Pinckney2023,
  author    = {Pinckney, Donald and Cassano, Federico and Guha, Arjun and Bell, Jonathan},
  title     = {A Large Scale Analysis of Semantic Versioning in NPM},
  booktitle = {Proceedings of the 20th IEEE/ACM International Conference on Mining Software Repositories (MSR 2023)},
  year      = {2023},
  pages     = {485--497},
  publisher = {IEEE},
  address   = {Piscataway, NJ, USA},
  doi       = {10.1109/MSR59073.2023.00073},
  isbn      = {9798350311846}
}

@article{Nocera2025,
author = {Nocera, Sabato and Penta, Massimiliano Di and Ahmed, Fatima and Romano, Simone and Scanniello, Giuseppe},
doi = {10.1145/3786773},
issn = {1049-331X},
journal = {ACM Trans. Softw. Eng. Methodol.},
number = {March},
pages = {1--37},
publisher = {ACM},
address   = {New York, NY, USA},
title = {{What We Know about AIBOMs: Results from a Multivocal Literature Review on Artificial Intelligence Bill of Materials}},
year = {2025}
}

@inproceedings{Costal2024,
author = {Costal, Dolors and G{\'{o}}mez, Cristina and {Del Rey}, Santiago and Mart{\'{i}}nez-Fern{\'{a}}ndez, Silverio},
doi = {10.1109/ICSA-C63560.2024.00055},
isbn = {9798350366259},
booktitle = {Proceedings of the IEEE 21st International Conference on Software Architecture Companion (ICSA-C 2024)},
publiser = {IEEE},
address = {Piscataway, NJ, USA},
pages = {289--294},
title = {{Using Metrics for Code Smells of ML Pipelines}},
year = {2024}
}

@article{Teyton2014,
  author    = {Teyton, C{\'{e}}dric and Falleri, Jean R{\'{e}}my and Palyart, Marc and Blanc, Xavier},
  title     = {A study of library migrations in Java},
  journal   = {Journal of Software: Evolution and Process},
  year      = {2014},
  volume    = {26},
  number    = {11},
  pages     = {1030--1052},
  publisher = {Wiley},
  address   = {Hoboken, NJ, USA},
  doi       = {10.1002/smr.1660}
}

@inproceedings{Gu2023,
author = {Gu, Haiqiao and He, Hao and Zhou, Minghui},
doi = {10.1109/SANER56733.2023.00064},
isbn = {9781665452786},
booktitle = {Proceedings of the 2023 IEEE International Conference on Software Analysis, Evolution and Reengineering (SANER 2023)},
pages = {627--638},
publisher = {IEEE},
address = {Piscataway, NJ, USA},
title = {{Self-Admitted Library Migrations in Java, JavaScript, and Python Packaging Ecosystems: A Comparative Study}},
year = {2023}
}

@inproceedings{He2021,
address = {New York, NY, USA},
author = {He, Hao and He, Runzhi and Gu, Haiqiao and Zhou, Minghui},
booktitle = {Proceedings of the 29th ACM Joint Meeting on European Software Engineering Conference and Symposium on the Foundations of Software Engineering (ESEC/FSE 2021)},
doi = {10.1145/3468264.3468571},
isbn = {9781450385626},
pages = {478--490},
publisher = {ACM},
title = {{A large-scale empirical study on Java library migrations: prevalence, trends, and rationales}},
year = {2021}
}

@misc{storey2026,
      title={From Technical Debt to Cognitive and Intent Debt: Rethinking Software Health in the Age of AI}, 
      author={Margaret-Anne Storey},
      year={2026},
      eprint={2603.22106},
      archivePrefix={arXiv},
      primaryClass={cs.SE},
      url={https://arxiv.org/abs/2603.22106}, 
}

@inbook{Jaisri2025,
address = {Cham, Switzerland},
author = {Jaisri, Pongchai and Reid, Brittany and Kula, Raula Gaikovina},
booktitle = {Software Engineering and Management: Theory and Applications: Volume 17},
doi = {10.1007/978-3-031-82610-8_4},
editor = {Lee, Roger},
isbn = {978-3-031-82610-8},
pages = {53--65},
publisher = {Springer Nature},
title = {{A Preliminary Study on Self-contained Libraries in the NPM Ecosystem}},
year = {2025}
}

@article{Islam2024,
  author = {Islam, Mohayeminul and Jha, Ajay Kumar and Akhmetov, Ildar and Nadi, Sarah},
  title = {Characterizing Python Library Migrations},
  journal   = {Proceedings of the ACM on Software Engineering},
  year = {2024},
  volume    = {1},
  number    = {FSE},
  pages     = {92--114},
  publisher = {ACM},
  address   = {New York, NY, USA},
  doi       = {10.1145/3643731}
}

@inproceedings{Geiping2023,
address = {Cambridge, MA, USA},
author = {Geiping, Jonas and others},
booktitle = {In Proceedings of the 40th International Conference on Machine Learning (ICML 2023)},
eprint = {2401.13177v1},
issn = {26403498},
pages = {11117--11143},
publisher = {PMLR Press},
title = {{Cramming: training a language model on a single GPU in one day}},
volume = {202},
month = {Jul},
year = {2023}
}

@inproceedings{Islam2023,
  author={Islam, Mohayeminul and Jha, Ajay Kumar and Nadi, Sarah and Akhmetov, Ildar},
  booktitle={Proceedings of the IEEE/ACM 20th International Conference on Mining Software Repositories (MSR 2023)}, 
  title={PyMigBench: A Benchmark for Python Library Migration}, 
  year={2023},
  publisher = {IEEE},
  address = {Piscataway, NJ, USA},
  pages={511-515},
  doi={10.1109/MSR59073.2023.00075}
}

@inproceedings{sd,
address = {Los Alamitos, CA, USA},
author = {Rombach, Robin and Blattmann, Andreas and Lorenz, Dominik and Esser, Patrick and Ommer, Bjorn},
booktitle = {Proceedings of the 2022 IEEE/CVF Conference on Computer Vision and Pattern Recognition (CVPR 2022)},
doi = {10.1109/CVPR52688.2022.01042},
month = {jun},
pages = {10674--10685},
publisher = {IEEE},
title = {{ High-Resolution Image Synthesis with Latent Diffusion Models }},
year = {2022}
}

@article{Han2021,
author = {Han, Xu and others},
issn = {2666-6510},
journal = {AI Open},
pages = {225--250},
title = {{Pre-trained models: Past, present and future}},
volume = {2},
publisher = {Elsevier},
address   = {Amsterdam, Netherlands},
doi = {10.1016/j.aiopen.2021.08.002},
year = {2021}
}

@inproceedings{Davis2023,
author = {Davis, James C. and others},
publisher = {IEEE},
address   = {Piscataway, NJ, USA},
doi = {10.1109/JVA60410.2023.00015},
isbn = {9798350328899},
booktitle = {Proceedings of the 2023 IEEE John Vincent Atanasoff Symposium on Modern Computing (JVA 2023)},
pages = {17--30},
title = {{Reusing Deep Learning Models: Challenges and Directions in Software Engineering}},
year = {2023}
}

@article{Ajibode2025,
author = {Ajibode, Adekunle and Bangash, Abdul Ali and Cogo, Filipe R. and Adams, Bram and Hassan, Ahmed E},
doi = {10.1007/s10664-025-10631-3},
isbn = {0123456789},
issn = {15737616},
publisher = {Springer},
address = {Cham, Switzerland},
journal = {Empirical Software Engineering},
number = {3},
pages = {1--63},
title = {{Towards semantic versioning of open pre-trained language model releases on hugging face}},
volume = {30},
year = {2025}
}

@article{Landis1977,
author = {Landis, J. R. and Koch, G. G.},
issn = {0006-341X (Print)},
journal = {Biometrics},
language = {eng},
month = {mar},
number = {1},
pages = {159--174},
pmid = {843571},
title = {{The measurement of observer agreement for categorical data}},
volume = {33},
publisher = {Wiley},
address   = {Hoboken, NJ, USA},
year = {1977}
}

@article{McHugh2012,
author = {McHugh, Mary L.},
issn = {1330-0962 (Print)},
journal = {Biochemia medica},
language = {eng},
number = {3},
pages = {276--282},
pmid = {23092060},
title = {{Interrater reliability: the kappa statistic}},
publisher = {Croatian Society of Medical Biochemistry and Laboratory Medicine},
address = {Zagreb, Croatia},
volume = {22},
year = {2012}
}

@inproceedings{Castano2024,
arxivId = {2311.13380},
author = {Castano, Joel and Martinez-Fernandez, Silverio and Franch, Xavier and Bogner, Justus},
booktitle = {Proceedings of the IEEE/ACM 21st International Conference on Mining Software Repositories (MSR 2024)},
doi = {10.1145/3643991.3644898},
eprint = {2311.13380},
isbn = {9798400705878},
number = {1},
pages = {607--618},
publisher = {ACM},
address   = {New York, NY, USA},
title = {{Analyzing the Evolution and Maintenance of ML Models on Hugging Face}},
volume = {1},
year = {2024}
}

@article{Jiang2025,
  author    = {Jiang, Wenxin and others},
  title     = {``I see models being a whole other thing'': an empirical study of pre-trained model naming conventions and a tool for enhancing naming consistency},
  journal   = {Empirical Software Engineering},
  year      = {2025},
  volume    = {30},
  number    = {6},
  pages     = {155},
  publisher = {Springer},
  address   = {Cham, Switzerland},
  doi       = {10.1007/s10664-025-10711-4}
}

@article{Bavota2015,
  author    = {Bavota, Gabriele and Canfora, Gerardo and {Di Penta}, Massimiliano and Oliveto, Rocco and Panichella, Sebastiano},
  title     = {How the Apache community upgrades dependencies: an evolutionary study},
  journal   = {Empirical Software Engineering},
  year      = {2015},
  volume    = {20},
  number    = {5},
  pages     = {1275--1317},
  publisher = {Springer},
  address   = {Cham, Switzerland},
  doi       = {10.1007/s10664-014-9325-9}
}

@article{Huang2022,
  author    = {Huang, Kaifeng and others},
  title     = {Characterizing usages, updates and risks of third-party libraries in Java projects},
  journal   = {Empirical Software Engineering},
  year      = {2022},
  volume    = {27},
  number    = {4},
  pages     = {78},
  publisher = {Springer},
  address   = {Cham, Switzerland},
  doi       = {10.1007/s10664-022-10131-8}
}

@inproceedings{Jiang2024,
author = {Jiang, Wenxin and others},
eprint = {2402.00699v1},
booktitle = {Proceedings of the 21st IEEE/ACM International Conference on Mining Software Repositories (MSR 2024)},
title = {{PeaTMOSS: A Dataset and Initial Analysis of Pre-Trained Models in Open-Source Software}},
volume = {1},
year = {2024},
doi = {10.1145/3643991.3644907},
publisher = {ACM},
pages = {431–443},
numpages = {13},
location = {Lisbon, Portugal},
address = {New York, NY, USA},
}

@inproceedings{Pepe2024,
author = {Pepe, Federica and others},
booktitle = {Proccedings of the 32nd IEEE International Conference on Program Comprehension (ICPC 2024)},
doi = {10.1145/3643916.3644412},
isbn = {9798400705861},
issn = {26437171},
number = {iii},
pages = {370--381},
publisher = {ACM},
address   = {New York, NY, USA},
title = {{How do Hugging Face Models Document Datasets, Bias, and Licenses? An Empirical Study}},
year = {2024}
}

@inproceedings{Chakraborty2021,
author = {Chakraborty, Mohna},
doi = {10.1145/3468264.3473494},
isbn = {9781450385626},
booktitle = {Proceedings of the 29th ACM Joint Meeting on European Software Engineering Conference and Symposium on the Foundations of Software Engineering (ESEC/FSE 2021)},
pages = {1686--1688},
title = {{Does reusing pre-trained NLP model propagate bugs?}},
publisher = {ACM},
year = {2021}
}

@inproceedings{Tan2024,
address = {Los Alamitos, CA, USA},
author = {Tan, Xin and others},
booktitle = {Proceedings of the 31st IEEE International Conference on Software Analysis, Evolution and Reengineering (SANER 2024)},
doi = {10.1109/SANER60148.2024.00015},
month = {mar},
pages = {67--78},
publisher = {IEEE},
title = {{Challenges of Using Pre-trained Models: the Practitioners' Perspective}},
year = {2024}
}

@inproceedings{Toma2024,
author = {Toma, Tajkia Rahman and Bezemer, Cor Paul},
booktitle = {Proceedings of the 3rd IEEE/ACM International Conference on AI Engineering -- Software Engineering for AI (CAIN 2024)},
doi = {10.1145/3644815.3644963},
isbn = {9798400705915},
pages = {64--74},
title = {{An exploratory study of dataset and model management in open source machine learning applications}},
year = {2024},
publisher = {ACM},
address   = {New York, NY, USA},
}

@inproceedings{Gonzalez2020,
author = {Gonzalez, Danielle and Zimmermann, Thomas and Nagappan, Nachiappan},
booktitle = {Proceedings of the 17th IEEE/ACM International Conference on Mining Software Repositories (MSR 2020)},
doi = {10.1145/3379597.3387473},
isbn = {9781450379571},
pages = {431--442},
title = {{The State of the ML-universe: 10 Years of Artificial Intelligence \& Machine Learning Software Development on GitHub}},
publisher = {ACM},
address = {New York, NY, USA},
year = {2020}
}

@article{Chakroborti2023,
  author    = {Chakroborti, Debasish and Nath, Sristy Sumana and Schneider, Kevin A. and Roy, Chanchal K.},
  title     = {Release conventions of open-source software: An exploratory study},
  journal   = {Journal of Software: Evolution and Process},
  year      = {2023},
  volume    = {35},
  number    = {1},
  pages     = {e2499},
  publisher = {Wiley},
  address   = {Hoboken, NJ, USA},
  doi       = {10.1002/smr.2499}
}

@inproceedings{Coelho2018,
author = {Coelho, Jailton and Valente, Marco Tulio and Silva, Luciana L. and Shihab, Emad},
booktitle = {Proceedings of the 12th ACM/IEEE International Symposium on Empirical Software Engineering and Measurement (ESEM 2018)},
publisher = {ACM},
address   = {New York, NY, USA},
doi = {10.1145/3239235.3240501},
eprint = {1809.04041},
isbn = {9781450358231},
issn = {19493789},
title = {{Identifying unmaintained projects in github}},
year = {2018}
}

@inproceedings{Joshi2019,
author = {Joshi, Saket Dattatray and Chimalakonda, Sridhar},
doi = {10.1109/MSR.2019.00088},
isbn = {9781728134123},
issn = {21601860},
booktitle = {Proceedings of the IEEE/ACM 16th International Conference on Mining Software Repositories (MSR 2019)},
pages = {587--591},
publisher = {IEEE},
address   = {Piscataway, NJ, USA},
title = {{RapidRelease - A dataset of projects and issues on github with rapid releases}},
volume = {2019-May},
year = {2019}
}

@inproceedings{Kilic2023,
author = {Kilic, Oz and Bowness, Nathaniel and Baysal, Olga},
doi = {10.1109/MSR59073.2023.00058},
isbn = {9798350311846},
booktitle = {Proceedings of the 20th IEEE/ACM International Conference on Mining Software Repositories (MSR 2023)},
pages = {372--376},
publisher = {IEEE},
address   = {Piscataway, NJ, USA},
title = {{Keep the Ball Rolling: Analyzing Release Cadence in GitHub Projects}},
year = {2023}
}

@misc{SemVer,
  author       = {T. Preston-Werner and Contributors},
  title        = {Semantic Versioning 2.0.0},
  howpublished = {\url{https://semver.org}},
  year         = {2023},
  note         = {Accessed: Oct. 13, 2025}
}

@article{Laukkanen2018,
  author    = {Laukkanen, Eero and Paasivaara, Maria and Itkonen, Juha and others},
  title     = {Comparison of release engineering practices in a large mature company and a startup},
  journal   = {Empirical Software Engineering},
  year      = {2018},
  volume    = {23},
  number    = {6},
  pages     = {3535--3577},
  publisher = {Springer},
  address   = {Cham, Switzerland},
  doi       = {10.1007/s10664-018-9616-7}
}

@article{Yeow2025,
author = {Yeow, Matthew Yit Hang and Chong, Chun Yong and Lim, Mei Kuan and {Yee Yen}, Yuen},
doi = {10.1371/journal.pone.0314512},
issn = {19326203},
journal = {PLoS ONE},
number = {2},
pages = {e0314512},
pmid = {39946354},
title = {{Predicting software reuse using machine learning techniques—A case study on open-source Java software systems}},
volume = {20},
month = {feb},
year = {2025},
publisher = {PLOS},
address   = {San Francisco, CA, USA},
}

@inproceedings{Taraghi2024,
archivePrefix = {arXiv},
arxivId = {2401.13177},
author = {Taraghi, Mina and Dorcelus, Gianolli and Foundjem, Armstrong and Tambon, Florian and Khomh, Foutse},
booktitle = {Proceedings of the 31st IEEE International Conference on Software Analysis, Evolution and Reengineering (SANER 2024)},
doi = {10.1109/SANER60148.2024.00059},
eprint = {2401.13177},
isbn = {9798350330663},
month = {mar},
pages = {512--523},
publisher = {IEEE},
address   = {Piscataway, NJ, USA},
title = {{Deep Learning Model Reuse in the HuggingFace Community: Challenges, Benefit and Trends}},
year = {2024}
}

@inproceedings{Jiang2022,
author = {Jiang, Wenxin and Synovic, Nicholas and Sethi, Rohan and others},
booktitle = {Proceedings of the 2022 ACM Workshop on Software Supply Chain Offensive Research and Ecosystem Defenses (SCORED 2022)},
doi = {10.1145/3560835.3564547},
isbn = {9781450398855},
pages = {105--114},
publisher = {ACM},
address   = {New York, NY, USA},
title = {{An Empirical Study of Artifacts and Security Risks in the Pre-trained Model Supply Chain}},
year = {2022}
}

@inproceedings{Yang2024,
  author    = {Yang, Xinyu and Liang, Weixin and Zou, James},
  title     = {Navigating Dataset Documentations in AI: A Large-Scale Analysis of Dataset Cards on Hugging Face},
  booktitle = {Proceedings of the 12th International Conference on Learning Representations (ICLR 2024)},
  year      = {2024},
  eprint    = {2401.13822},
  archivePrefix = {arXiv},
  publisher = {OpenReview.net}
}

@inproceedings{Jones2024,
  author    = {Jones, Jason and others},
  title     = {What do we know about Hugging Face? A systematic literature review and quantitative validation of qualitative claims},
  booktitle = {Proceedings of the 18th ACM/IEEE International Symposium on Empirical Software Engineering and Measurement (ESEM 2024)},
  year      = {2024},
  number = {1},
  volume = {1},
  pages     = {13--24},
  publisher = {ACM},
    address   = {New York, NY, USA},
  doi       = {10.1145/3674805.3686665},
  eprint    = {2406.08205},
  archivePrefix = {arXiv},
  isbn = {9798400710476},
issn = {19493789},
}

@incollection{Woolson2008,
  author    = {Woolson, R. F.},
  title     = {Wilcoxon Signed-Rank Test},
  booktitle = {Wiley Encyclopedia of Clinical Trials},
  publisher = {John Wiley \& Sons, Ltd},
  year      = {2008},
  pages     = {1--3},
  doi       = {10.1002/9780471462422.eoct979}
}

@inproceedings{Castano2023,
author = {Castano, Joel and Martinez-Fernandez, Silverio and Franch, Xavier and Bogner, Justus},
booktitle = {Proceedings of the 17th IEEE/ACM International Symposium on Empirical Software Engineering and Measurement (ESEM 2023)},
doi = {10.1109/ESEM56168.2023.10304801},
eprint = {2305.11164},
isbn = {9781665452236},
issn = {19493789},
pages = {1--12},
publisher = {IEEE},
address   = {Piscataway, NJ, USA},
title = {{Exploring the Carbon Footprint of Hugging Face's ML Models: A Repository Mining Study}},
year = {2023}
}

@misc{phi4,
      title={Phi-4-reasoning Technical Report}, 
      author={Marah Abdin and Sahaj Agarwal and Ahmed Awadallah and others},
      year={2025},
      eprint={2504.21318},
      archivePrefix={arXiv},
      primaryClass={cs.AI},
      url={https://arxiv.org/abs/2504.21318}, 
}

@incollection{Diener2010,
  author    = {Diener, Marc J.},
  title     = {Cohen's d},
  booktitle = {The Corsini Encyclopedia of Psychology},
  publisher = {John Wiley \& Sons, Ltd},
  year      = {2010},
  pages     = {1},
  doi       = {10.1002/9780470479216.corpsy0200}
}

@article{Mann1947,
  author    = {Mann, H. B. and Whitney, D. R.},
  title     = {On a Test of Whether One of Two Random Variables Is Stochastically Larger than the Other},
  journal   = {The Annals of Mathematical Statistics},
  year      = {1947},
  volume    = {18},
  number    = {1},
  pages     = {50--60},
  publisher = {Institute of Mathematical Statistics},
  address   = {Beachwood, OH, USA},
  doi       = {10.1214/aoms/1177730491}
}

@inproceedings{Montes2022,
author = {Montes, Diego and Peerapatanapokin, Pongpatapee and Schultz, Jeff and Guo, Chengjun and Jiang, Wenxin and Davis, James C.},
booktitle = {Proceedings of the 30th ACM Joint Meeting on European Software Engineering Conference and Symposium on the Foundations of Software Engineering (ESEC/FSE 2022)},
doi = {10.1145/3540250.3560881},
isbn = {9781450394130},
pages = {1605--1609},
publisher = {ACM},
address   = {New York, NY, USA},
title = {{Discrepancies among pre-trained deep neural networks: A new threat to model zoo reliability}},
year = {2022}
}

@article{Gong2023,
author = {Gong, Lina and Zhang, Jingxuan and Wei, Mingqiang and Zhang, Haoxiang and Huang, Zhiqiu},
doi = {10.1145/3569934},
issn = {15577392},
journal = {ACM Trans. Softw. Eng. Methodol.},
number = {3},
title = {{What Is the Intended Usage Context of This Model? An Exploratory Study of Pre-Trained Models on Various Model Repositories}},
volume = {32},
year = {2023},
month = {may},
pages = {1--57},
articleno = {69},
numpages = {57},
publisher = {ACM},
address   = {New York, NY, USA},
}

@inproceedings{banyongrakkul2025,
author={Banyongrakkul, Peerachai and Zahedi, Mansooreh and Thongtanunam, Patanamon and Treude, Christoph and Gao, Haoyu},
booktitle = {Proceedings of the 41st IEEE International Conference on Software Maintenance and Evolution (ICSME 2025)}, 
title={From Release to Adoption: Challenges in Reusing Pre-Trained AI Models for Downstream Developers}, 
year={2025},
pages={1--13},
publisher = {IEEE},
address   = {Piscataway, NJ, USA},
doi={10.1109/ICSME64153.2025.00022}
}

@article{Yasmin2025,
arxivId = {arXiv:2509.06085v1},
year={2026},
author = {Yasmin, Jerin and Jiang, Wenxin and Tian, C Davis Yuan},
eprint = {arXiv:2509.06085v1},
journal = {},
eprint={2509.06085},
archivePrefix={arXiv},
primaryClass={cs.SE},
title = {{Software Dependencies 2.0: An Empirical Study of Reuse and Integration of Pre-Trained Models in Open-Source Projects}}
}

@article{Gao2025,
eprint={2503.19444},
archivePrefix={arXiv},
primaryClass={cs.SE},
author = {Gao, Haoyu and Zahedi, Mansooreh and Jiang, Wenxin and Lin, Hong Yi and Davis, James and Treude, Christoph},
eprint = {arXiv:2503.19444v3},
number = {1},
pages = {1--29},
title = {{AI Safety in the Eyes of the Downstream Developer: A First Look at Concerns, Practices, and Challenges}},
volume = {1},
year = {2025},
journal = {}
}

@article{Ratner2002,
author = {Ratner, Carl},
issn = {14385627},
journal = {Forum Qualitative Sozialforschung},
number = {3},
title = {{Subjectivity and objectivity in qualitative methodology}},
volume = {3},
year = {2002},
pages     = {Art. 16},
publisher = {Free University of Berlin},
address = {Berlin, Germany}
}

@article{Gao2026,
archivePrefix = {arXiv},
arxivId = {2603.00489},
author = {Gao, Haoyu and Lin, Hong Yi and Treude, Christoph and Gay, Gregory and Zahedi, Mansooreh},
eprint = {2603.00489},
pages = {1--18},
title = {{Does My README File Need To Be Updated? Exploring LLM-Based README Maintenance}},
year = {2026},
journal = {}
}

@article{Leipzig2021,
author = {Leipzig, Jeremy and N{\"{u}}st, Daniel and Hoyt, Charles Tapley and Ram, Karthik and Greenberg, Jane},
doi = {10.1016/j.patter.2021.100322},
issn = {2666-3899 (Electronic)},
journal = {Patterns},
address = {New York, NY, USA},
language = {eng},
month = {sep},
number = {9},
pages = {100322},
pmid = {34553169},
title = {The Role of Metadata in Reproducible Computational Research},
volume = {2},
year = {2021},
publisher = {Elsevier}
}

@article{Soto-Valero2021,
author = {Soto-Valero, C{\'{e}}sar and Harrand, Nicolas and Monperrus, Martin and Baudry, Benoit},
doi = {10.1007/s10664-020-09914-8},
issn = {1573-7616},
journal = {Empirical Software Engineering},
publisher = {Springer},
address   = {Cham, Switzerland},
number = {3},
pages = {45},
title = {{A comprehensive study of bloated dependencies in the Maven ecosystem}},
volume = {26},
year = {2021}
}

@inproceedings{Zaimi2015,
address = {New York, NY, USA},
author = {Zaimi, Asimina and others},
booktitle = {Proceedings of the 7th Balkan Conference on Informatics Conference (BCI 2015)},
doi = {10.1145/2801081.2801087},
isbn = {9781450333351},
publisher = {ACM},
title = {{An Empirical Study on the Reuse of Third-Party Libraries in Open-Source Software Development}},
year = {2015}
}

@inproceedings{Terzi2022,
author = {Terzi, Anastasia},
doi = {10.1109/SEAA56994.2022.00048},
isbn = {9781665461528},
booktitle = {Proceedings of the 48th Euromicro Conference on Software Engineering and Advanced Applications (SEAA 2022)},
pages = {263--269},
publisher = {IEEE},
title = {{Software Reuse and Evolution in JavaScript Applications}},
year = {2022}
}

@book{Gao2026b,
archivePrefix = {arXiv},
arxivId = {2601.13754},
author = {Gao, Haoyu and Banyongrakkul, Peerachai and Guan, Hao and Zahedi, Mansooreh and Treude, Christoph},
booktitle = {Proceedings of the 23rd International Conference on Mining Software Repositories (MSR 2026)},
eprint = {2601.13754},
publisher = {IEEE},
address = {Los Alamitos, CA, USA},
title = {{On Autopilot? An Empirical Study of Human-AI Teaming and Review Practices in Open Source}},
year = {2026}
}

@incollection{Gamma1993,
  author    = {Gamma, Erich and Helm, Richard and Johnson, Ralph and Vlissides, John},
  title     = {Design Patterns: Abstraction and Reuse of Object-Oriented Design},
  booktitle = {Proceedings of the European Conference on Object-Oriented Programming (ECOOP 1993)},
  series    = {Lecture Notes in Computer Science},
  volume    = {707},
  pages     = {406--431},
  year      = {1993},
  publisher = {Springer Nature},
  address = {Cham, Switzerland},
  doi       = {10.1007/3-540-47910-4_21}
}

@article{Giordano2023,
author = {Giordano, Giammaria and others},
doi = {10.1007/s10664-023-10408-6},
issn = {1573-7616},
journal = {Empirical Software Engineering},
number = {1},
pages = {20},
title = {{On the adoption and effects of source code reuse on defect proneness and maintenance effort}},
publisher = {Springer},
address   = {Cham, Switzerland},
volume = {29},
year = {2023}
}

@article{Zhang2022,
author = {Zhang, Jie M and Harman, Mark and Ma, Lei and Liu, Yang},
doi = {10.1109/TSE.2019.2962027},
journal = {IEEE Trans. Softw. Eng.},
publisher = {IEEE},
address   = {Piscataway, NJ, USA},
number = {1},
pages = {1--36},
title = {{Machine Learning Testing: Survey, Landscapes and Horizons}},
volume = {48},
year = {2022}
}

@article{Nocera2025b,
author = {Nocera, Sabato and others},
doi = {10.1016/j.jss.2025.112540},
issn = {01641212},
journal = {Journal of Systems and Software},
number = {July},
pages = {112540},
publisher = {Elsevier},
address   = {Amsterdam, Netherlands},
title = {{On the adoption of software bill of materials in open-source software projects}},
volume = {230},
year = {2025}
}

@inproceedings{Gesese2025,
author = {Gesese, Genet Asefa and Chen, Zongxiong and Zoubia, Oussama and others},
isbn = {0000000279307},
issn = {16130073},
booktitle = {Proceedings of the CEUR Workshop},
pages = {57--71},
title = {{A Survey on Metadata for Machine Learning Models and Datasets: Standards, Practices, and Harmonization Challenges}},
volume = {4065},
year = {2025},
publisher = {CEUR-WS.org}
}

@article{thomas2006,
author = {Thomas, David R},
doi = {10.1177/1098214005283748},
journal = {American Journal of Evaluation},
number = {2},
pages = {237--246},
title = {{A General Inductive Approach for Analyzing Qualitative Evaluation Data}},
volume = {27},
year = {2006},
publisher = {SAGE Publications},
address = {Thousand Oaks, CA, USA}
}

@inproceedings{Barbosa2022,
  author    = {Barbosa, L{\'{i}}via and Hora, Andre},
  title     = {How and Why Developers Migrate Python Tests},
  booktitle = {Proceedings of the IEEE International Conference on Software Analysis, Evolution and Reengineering (SANER 2022)},
  year      = {2022},
  pages     = {538--548},
  publisher = {IEEE},
  address   = {Piscataway, NJ, USA},
  doi       = {10.1109/SANER53432.2022.00071}
}

@inproceedings{Kabinna2016,
author = {Kabinna, Suhas and Bezemer, Cor-Paul and Shang, Weiyi and Hassan, Ahmed E},
booktitle = {Proceedings of the 13th International Conference on Mining Software Repositories (MSR 2016)},
doi = {10.1145/2901739.2901769},
isbn = {9781450341868},
pages = {154--164},
publisher = {ACM},
title = {{Logging library migrations: a case study for the apache software foundation projects}},
year = {2016}
}

@online{commit_swin_addition,
  author = {ZFTurbo/Music-Source-Separation-Training},
  title = {Commit 7ee8e07},
  year = {2023},
  url = {https://github.com/ZFTurbo/Music-Source-Separation-Training/commit/7ee8e074e6a9f6cd217f66a360a82c84cc2b174a},
  note = {(Accessed: 2025-12-01)}
}

@online{commit_roberta_removal,
  author = {foundation-model-stack/aiu-fms-testing-utils},
  title = {Commit 041f39a},
  year = {2025},
  url = {https://github.com/foundation-model-stack/aiu-fms-testing-utils/commit/041f39a9},
  note = {(Accessed: 2025-12-01)}
}

@online{example1,
  author = {luxonis/datadreamer},
  title = {Example 1},
  year = {2025},
  url = {https://github.com/luxonis/datadreamer/pull/77},
  note = {(Accessed: 2025-12-01)}
}

@online{example2,
  author = {centre-for-humanities-computing/conspiracies},
  title = {Example 2},
  year = {2023},
  url = {https://github.com/centre-for-humanities-computing/conspiracies/commit/2c3d5e32318dd0713770d32b485b63ff986e67ac},
  note = {(Accessed: 2025-12-01)}
}

@online{example3,
  author = {hpcaitech/ColossalAI},
  title = {Example 3},
  year = {2024},
  url = {https://github.com/hpcaitech/ColossalAI/releases/tag/v0.4.3},
  note = {(Accessed: 2025-12-01)}
}

@online{example4,
  author = {koito19960406/ZenSVI},
  title = {Example 4},
  year = {2024},
  url = {https://github.com/koito19960406/ZenSVI/pull/91},
  note = {(Accessed: 2025-12-01)}
}

@online{example5,
  author = {luxonis/datadreamer},
  title = {Example 5},
  year = {2024},
  url = {https://github.com/vllm-project/vllm/issues/4141},
  note = {(Accessed: 2025-12-01)}
}

@online{example6,
  author = {castorini/pyserini},
  title = {Example 6},
  year = {2021},
  url = {https://github.com/castorini/pyserini/pull/620},
  note = {(Accessed: 2025-12-01)}
}

@online{example7,
  author = {huggingface/trl},
  title = {Example 7},
  year = {2025},
  url = {https://github.com/huggingface/trl/pull/3415},
  note = {(Accessed: 2025-12-01)}
}

@online{example8,
  author = {biopragmatics/bioregistry},
  title = {Example 8},
  year = {2025},
  url = {https://github.com/biopragmatics/bioregistry/pull/1439/commits/e29600af8d57c7dacf28d7bddddb3b629f2e0b1a},
  note = {(Accessed: 2025-12-01)}
}

@online{example9,
  author = {luxonis/datadreamer},
  title = {Example 9},
  year = {2024},
  url = {https://github.com/TransformerLensOrg/TransformerLens/pull/777},
  note = {(Accessed: 2025-12-01)}
}

@online{example10,
  author = {vllm-project/vllm},
  title = {Example 10},
  year = {2025},
  url = {https://github.com/vllm-project/vllm/pull/14422},
  note = {(Accessed: 2025-12-01)}
}

@online{example11,
  author = {PrunaAI/pruna},
  title = {Example 11},
  year = {2025},
  url = {https://github.com/PrunaAI/pruna/commit/bc1ece9b77f4fd426fbaf43e03b2f5eb66f2dc96},
  note = {(Accessed: 2025-12-01)}
}

@online{example12,
  author = {arthur-ai/arthur-engine},
  title = {Example 12},
  year = {2025},
  url = {https://github.com/arthur-ai/arthur-engine/pull/310},
  note = {(Accessed: 2025-12-01)}
}

@online{example13,
  author = {mlflow/mlflow},
  title = {Example 13},
  year = {2023},
  url = {https://github.com/mlflow/mlflow/pull/8623},
  note = {(Accessed: 2025-12-01)}
}

@online{example14,
  author = {mlflow/mlflow},
  title = {Example 14},
  year = {2024},
  url = {https://github.com/mlflow/mlflow/issues/10887},
  note = {(Accessed: 2025-12-01)}
}

@online{example15,
  author = {vllm-project/vllm},
  title = {Example 14},
  year = {2025},
  url = {https://github.com/vllm-project/vllm/pull/21169},
  note = {(Accessed: 2025-12-01)}
}

@online{example16,
  author = {MannLabs/scPortrait},
  title = {Example 16},
  year = {2025},
  url = {https://github.com/MannLabs/scPortrait/issues/308},
  note = {(Accessed: 2025-12-01)}
}

@online{example17,
  author = {mlflow/mlflow},
  title = {Example 17},
  year = {2024},
  url = {https://github.com/mlflow/mlflow/pull/11893},
  note = {(Accessed: 2025-12-01)}
}

@online{example18,
  author = {astorini/pyserini},
  title = {Example 18},
  year = {2022},
  url = {https://github.com/castorini/pyserini/pull/1113},
  note = {(Accessed: 2025-12-01)}
}

@online{example19,
  author = {astorini/pyserini},
  title = {Example 19},
  year = {2021},
  url = {https://github.com/castorini/pyserini/commit/af334c111a2d19f5331f060e88c6eae0267165f1},
  note = {(Accessed: 2025-12-01)}
}

@online{example20,
  author = {nunchaku-ai/nunchaku},
  title = {Example 20},
  year = {2025},
  url = {https://github.com/nunchaku-ai/nunchaku/commit/6dc8e717ae183e284197e77cca9dd2d13b3209d5#diff-40019c200c322fe31fa5dba87e1af010a4692c92b90cb591d142cda0b774cbe5},
  note = {(Accessed: 2025-12-01)}
}

@online{meleksabit2025example,
  author = {meleksabit/ai-powered-alerting-system},
  title = {Example 21},
  year = {2025},
  url = {https://github.com/meleksabit/ai-powered-alerting-system/blob/1fb500b46e5194e4973a32108b44acb9f2102e73/my_app/app.py},
  note = {(Accessed: 2025-12-01)}
}

\end{document}